\documentclass[journal,draftcls,onecolumn,twoside,11pt]{IEEEtran}
\normalsize
\ifCLASSINFOpdf
\else
\fi

\usepackage{amsmath}
\usepackage{cite}
\usepackage{graphicx}
\usepackage{wrapfig}
\usepackage{caption}
\usepackage{subcaption}
\usepackage{float}
\usepackage{setspace}
\usepackage{multirow}

\restylefloat{figure}
\input{epsf}
\begin{document}
\doublespacing

\title{Information Rates of ASK-Based Molecular Communication in Fluid Media}

\author{Siavash~Ghavami,~\IEEEmembership{Student~Member,~IEEE,}
        Raviraj~Adve,~\IEEEmembership{Senior~Member,~IEEE}
        and~Farshad~Lahouti,~\IEEEmembership{Senior~Member,~IEEE}% <-this % stops a space
\thanks{S. Ghavami and F. Lahouti are with the School
of E\&CE, College of Engineering, University of Tehran, e-mail: (s.ghavami,lahouti)@ut.ac.ir. F. Lahouti is a visiting professor of electrical engineering at California Institute of Technology, email: lahouti@caltech.edu.}% <-this % stops a space
\thanks{R. Adve is with the Department of E\&CE,
University of Toronto, email: rsadve@ece.utoronto.ca.}% <-this % stops a space
%\thanks{Mr.~Ghavami was supported in part by the Iranian Ministry of Science, Research and Technology  and in part by the National Science and Engineering Research Council (NSERC) Canada.}
}

%\markboth{IEEE Trans. on Molecular, Biological, and Multi-Scale Communications}%
%{Submitted paper}

\maketitle
%\vspace*{-1.in}
\begin{abstract}
%\boldmath
This paper studies the capacity of molecular communications in fluid media, where the information is encoded in the number of transmitted molecules in a time-slot (amplitude shift keying). The propagation of molecules is governed by random Brownian motion and the communication is in general subject to inter-symbol interference (ISI). We first consider the case where ISI is negligible and analyze the capacity and the capacity per unit cost of the resulting discrete memoryless molecular channel and the effect of possible practical constraints, such as limitations on peak and/or average number of transmitted molecules per transmission. In the case with a constrained peak molecular emission, we show that as the time-slot duration increases, the input distribution achieving the capacity per channel use transitions from binary inputs to a discrete uniform distribution. In this paper, we also analyze the impact of ISI. Crucially, we account for the correlation that ISI induces between channel output symbols. We derive an upper bound and two lower bounds on the capacity in this setting. Using the input distribution obtained by an extended Blahut-Arimoto algorithm, we maximize the lower bounds. Our results show that, over a wide range of parameter values, the bounds are close.
\end{abstract}

\vspace*{-0.25in}

\IEEEpeerreviewmaketitle

\section{Introduction}
As nano-technology has received increasing attention, researchers have investigated communications based on the release of molecules in a fluid propagation medium and their detection at a receiver. Diffusion-based channels are of particular interest because diffusion is the basis of data transmission in most of cell signaling, e.g., calcium signaling, hormones, etc.~\cite{20}. Diffusion-based molecular communications systems encode information in the concentration~\cite{1}, time of release~\cite{2}, the number of molecules released in a time-slot (amplitude-shift keying (ASK)), the type of molecules and the ratio of different types of molecules~\cite{Chae}. The released molecules propagate in a fluid medium before arriving at the receiver. Of these choices, this paper focuses on amplitude shift keying and characterizes the capacity of a diffusion molecular communication channel in different settings.

The movement of molecules in a fluid medium is governed by Brownian motion~\cite{3}, including when the medium itself moves with a drift velocity~\cite{MCBook}. The propagation time between transmitter and receiver is, therefore, random. Several works have investigated the capacity of molecular communications using a fluid medium.  In~\cite{21}, the authors study the capacity of a discrete molecular diffusion-based channel when the information is encoded in the  concentration of molecules. The work described in~\cite{22}, analyzes the capacity of molecular communications with on-off keying. While most of  the initial works assumed perfect synchronization, the work in~\cite{25} arrives at a more realistic model considering sequences of molecules. Recent results have also considered noise, memory and the impact of the associated physical properties in diffusion-based communications~\cite{26}.  In  addition  to  capacity,  modulation, error  probability, and symbol interval optimization have been investigated in, e.g.,~\cite{27}~\cite{28}. The capacity of molecular communication when information is encoded in the concentration of molecules released is studied for binary communications in~\cite{5,6} and for binary and 4-ary communications in~\cite{7}. In\cite{5,6,7}, the aggregate distribution of the number of arrived molecules in a time-slot is approximated by a Gaussian distribution.

In~\cite{2}, the random propagation time between transmitter and receiver was shown to follow an additive inverse Gaussian (AIG) distribution for 1-dimensional propagation. Based on the AIG distribution, expressions and bounds for channel capacity are presented in ~\cite{2, Guo, EckPeak}, for the case when information is encoded in the release time of molecules.

In the model considered here, in which data transfer is based on ASK, the transmitter releases a chosen number of molecules into a 1-D fluid medium with drift. The molecules then propagate to the receiver where they are detected and removed from the system. One of the main challenges in such a diffusion-based system is inter-symbol interference (ISI): due to the random propagation time, molecules may arrive over many time-slots. Indeed, if the system did not suffer from ISI, in our model wherein molecule release and detection are perfect, communications would also be perfect. The work in~\cite{3}, disregarding ISI, studies a binary ASK scheme by considering the life expectancy of molecule with an AIG model for propagation time. Of interest here is a capacity analysis of an ASK molecular system suffering from the effects of ISI. In~\cite{31}, an achievable rate and the probability of error of binary and 4-ary ASK molecular schemes are investigated in presence of ISI. Specifically, the analysis there assumes that output symbols are mutually independent, resulting in a lower bound on capacity~\cite{10}.

Capacity analyses of conventional communication channels with ISI has a strong presence in the literature. Several methods have been proposed, starting from the seminal work by Hirt and Massey~\cite{8}, which analyzes the capacity of the Gaussian ISI channel. The problem is solved for the case of an unrestricted input distribution, but remains open for a discrete input alphabet~\cite{9}. However, several bounds, both lower and upper, to the capacity with discrete inputs and Gaussian channels have been obtained~\cite{10}.  A simulation-based approach to calculate capacity in the case of i.i.d. (independent and identically distributed)~inputs with binary modulation is given in~\cite{14}. This approach is based on a trellis structure where the number of states relates to the memory of the ISI channel.%~\cite{12,13,14,15}. \cite{10,11}

In this paper, with information encoded in the number of molecules released, the receiver counts the number of molecules received within each time-slot. We obtain the probability of molecules arriving within a specific time-slot using the AIG distribution; this leads to a binomial distribution on the number of molecules received in each time-slot. We begin by analyzing ASK over such a channel in the basic discrete memoryless setting. In Section III we model the system as a memoryless channel. This molecular discrete memoryless channel (DMC) may be motivated by a molecular communication system with life expectancy of molecules being equal to the time-slot duration. This implies that the molecules disappear or decompose after the duration of a time slot. Specifically, in Section III, the molecules do not cause ISI.

We also study the capacity per channel use, the capacity per unit time, and the capacity per unit cost and examine the effects of possible constraints on peak and average molecular cost. Capacity per channel use and time are useful measures in applications such as drug delivery systems (DDS)~\cite{Akyl} and communication over fluid channels. The capacity per unit cost is the objective function in cost efficient communications, where the information rate per molecular emission is more important. To the best of our knowledge, this is the first study of capacity per unit cost in molecular communications. Constraints on transmission cost address certain practical constraints in molecular communications. Specifically, the peak constraint acts as a toxicity constraint in DDS~\cite{Akyl}. Similarly, the average constraint reflects a limitation on the average input emission or injection rate in molecular communications or DDS~\cite{Alen}. For DMC, we obtain the optimal input distribution and the resulting quantity for different capacity measures with average or peak molecular cost constraints. We also use this analysis to obtain the optimum transmission symbol interval as a function of the channel model parameters in each case.

We then move on to analyzing the impact of ISI in ASK-based molecular communications. As in~\cite{10}, with some abuse of notation, we call $I_{i.i.d}$ the capacity with i.i.d.~inputs, the channel capacity. We propose two lower bounds and an upper bound for the $I_{i.i.d}$ of such a channel. To provide a tractable exposition, we restrict the effect of ISI to one time-slot and analyze the average mutual information with correlated channel outputs. The lower bounds are maximized by optimizing the input distribution using a modified Blahut-Arimoto algorithm. As our results show, over a large range of parameter values, the lower and upper bounds are close, thereby well characterizing the capacity of such a molecular communication channel.

A crucial question left unanswered by the presented capacity analysis is what the impact of the simplifying assumption on one time-slot memory is. We are unable to directly evaluate the quality of this assumption since the true channel capacity is unknown. As a proxy, we develop the maximum likelihood (ML) detector as a possible implementation of the molecular communication receiver. We evaluate the error rates assuming ISI restricted to one time-slot and compare the performance to simulations which do not impose this assumption. As we will see, this allows us to characterize the range of parameters over which the assumption holds true.

The remainder of this paper is organized as follows: the system model and the problem statement are presented in Section~\ref{sec:model}. Section~\ref{sec:DMC} focuses on capacity analyses of a discrete memoryless ASK molecular communications, while Section~\ref{sec:ISI} presents our analysis in presence of ISI. Numerical results illustrating the analyses are presented in each section. Finally, Section V wraps up the paper with concluding remarks.

\section{System Model and Problem Statement} \label{sec:model}
\subsection{System Model}
The transmitter is a point source of identical molecules. Using ASK, at the beginning of every time-slot of length $T$, it transmits a message by releasing $X$, $0 \le X \le {X_{\max }}$, molecules into the fluid medium. Here, $X_{\max}$ is the maximum number of molecules releasable in any time-slot. The transmitter does not affect the propagation of the molecules. The channel is one-dimensional. Molecules propagate between the transmitter and receiver by Brownian motion characterized by a diffusion constant $d$ and (positive) drift velocity $v$.

At the receiver, all received molecules are absorbed and removed from the system. We consider two cases: first, we ignore ISI and assume that a molecule released in time-slot $m$ arrives within the same time-slot or disappears; this leads to a DMC. In the second half of the paper, we consider one slot of ISI, i.e., we assume molecules that do not arrive within two time-slots have disappeared\footnote{Note that one could use our analysis to avoid this assumption. However, the exposition becomes unwieldy and exponentially complicated to deal with.}. Everything else within the system operates perfectly, i.e., the only randomness in our model is the propagation time (equivalently, the number of molecules that are received in a time-slot). If $l$ denotes the distance between transmitter and receiver \footnote{The units of $l$ and $v$ are normally in $\mu m$ and $\mu m/sec$ but but any scaled version of these units can be used}, using the AIG analysis in~\cite{2}, the cumulative distribution function (CDF) of the propagation time is given by
\begin{equation}\label{eq:1}
{F_W}\left( w \right) = \Phi \left( {\sqrt {\frac{\lambda }{w}} \left( {\frac{w}{\mu } - 1} \right)} \right) + {e^{\frac{{2\lambda }}{\mu }}}\Phi \left( { - \sqrt {\frac{\lambda }{w}} \left( {\frac{w}{\mu } + 1} \right)} \right),w > 0.
\end{equation}
Here, $\Phi \left(  \cdot  \right)$ is the CDF of a standard Gaussian random variable, $\mu = l/v$, $\lambda = l^2/\sigma^2$ and $\sigma^2 = d/2$ is the variance of the associated Weiner process~\cite{2}. It is noteworthy that in a 3-dimensional environment the first passage process is transient, i.e., there is a non-zero probability of the molecule never arriving at the receiver~\cite{Ziff}. On the other hand, this is not true for the 1-D propagation under consideration; the first passage process for the 1-D case is said to be ``recurrent".

When the transmitter releases $X = x$ molecules in a time-slot, the probability of receiving $Y = y$ of these molecules in the same time-slot is given by
\begin{equation} \label{eq:2}
\Pr \left( {\left. {{Y} = y} \right|{X} = x} \right) = \left\{ {\begin{array}{*{20}{c}}
{\left( {\begin{array}{*{20}{c}}
x\\
y
\end{array}} \right)q_1^y{{\left( {1 - {q_1}} \right)}^{x - y}}{\rm{ }},{\rm{0}} \le y \le x}\\
{\begin{array}{*{20}{c}}
0,&{}&{}&{}&{}
\end{array}y < 0,y > x},
\end{array}} \right.
\end{equation}
where $q_1 = F_W(T)$ is the  probability of a molecule arriving within the same time-slot of its release. In general, $q_k = F_W(kT)-F_W((k-1)T)$ denotes the probability of a molecule arriving in the $k$-th time-slot after transmission.

At the start of time-slot $k$, ${X_k} \in \left\{ {0,1,...,{X_{\max }}} \right\}$  molecules are released. Let $a_x = \Pr\left(X = x\right)$, $0 \le x \le {X_{\max }}$, denote the probability of releasing $x$ molecules at the transmitter, and $Z_k$ denote the number of molecules in time-slot $k$, that were released but \emph{not received} in slot $k$. If $Y_k$ denotes the number of molecules received in time-slot $k$, we have
\begin{equation}\label{eq:5}
Y_k = X_k - Z_k + N_{k-1},\hspace{1cm}{Y_k} \in \left\{ {0,1,...,k{X_{\max }}} \right\},
\end{equation}
in which, $N_{k-1}\in \left\{ {0,1,...,{(k - 1) X_{\max }}} \right\}$ denotes the total number of ``interfering" molecules from the previous $k-1$ time-slots arriving in time-slot $k$. Considering ${P_k}\left( n \right) = \Pr(N_{k-1}=n)$, the channel transition probability when ${X_k} = x $ molecules are transmitted is given by
\begin{equation}\label{eq:6}
\begin{array}{l}
{p_{{{Y_k}} \mid {X_k}}}\left( {y \mid X_k = x} \right) = {\left( {1 - {q_1}} \right)^x}{P_k}\left( y \right) + \left( {\begin{array}{*{20}{c}}
x\\
1
\end{array}} \right){q_1}{\left( {1 - {q_1}} \right)^{x - 1}}{P_k}\left( {y - 1} \right) + ...\\
\left( {\begin{array}{*{20}{c}}
x\\
{x - 1}
\end{array}} \right)q_1^{x - 1}\left( {1 - q} \right){P_k}\left( {y - \left( {x - 1} \right)} \right) + q_1^x{P_k}\left( {y - x} \right),{\rm{   }} \hspace*{0.3in} y = 0,...,{X_{\max }}k - \left( {{X_{\max }} - x} \right).
\end{array}
\end{equation}
In \eqref{eq:6}, $P_k(y)$ can be easily calculated by induction as
\begin{equation}\label{eq:7}
\begin{array}{l}
{P_k}\left( y \right) = \left( {{a_0} + \left( {1 - {q_k}} \right){a_1} + ... + {{\left( {1 - {q_k}} \right)}^{{X_{\max }}}}{a_{{X_{\max }}}}} \right){P_{k - 1}}\left( y \right) + \\
\left( {{q_k}{a_1} + \left( {\begin{array}{*{20}{c}}
2\\
1
\end{array}} \right){q_k}\left( {1 - {q_k}} \right){a_2}{\rm{ + }}...{\rm{ + }}\left( {\begin{array}{*{20}{c}}
{{X_{\max }}}\\
1
\end{array}} \right){q_k}{{\left( {1 - {q_k}} \right)}^{{X_{\max }} - 1}}{a_{{X_{\max }}}}} \right){P_{k - 1}}\left( {y - 1} \right) + \\
 \vdots \\
\left( {q_k^{{X_{\max }} - 1}{a_{{X_{\max }} - 1}} + \left( {\begin{array}{*{20}{c}}
{{X_{\max }}}\\
{{X_{\max }} - 1}
\end{array}} \right)q_k^{{X_{\max }} - 1}\left( {1 - {q_k}} \right){a_{{X_{\max }}}}} \right){P_{k - 1}}\left( {y - \left( {{X_{\max }} - 1} \right)} \right) + \\
q_k^{{X_{\max }}}{a_{{X_{\max }}}}{P_{k - 1}}\left( {y - {X_{\max }}} \right)
\end{array}
\end{equation}
where, as defined, ${a_i} = \Pr\left( {{X} = i} \right)$  for $i \in \left\{ {0,...,{X_{\max }}} \right\}$. If, as later in Section~\ref{sec:ISI}, we were to assume that the ISI only affects the next time-slot,~\eqref{eq:7} is simplified to
\begin{subequations}\label{eq:9B}
\begin{align}
\label{eq:9Ba}
&{P_2}\left( 0 \right) = \left( {{a_0} + \left( {1 - {q_2}} \right){a_1} + ... + {{\left( {1 - {q_2}} \right)}^{{X_{\max }}}}{a_{{X_{\max }}}}} \right){P_1}\left( 0 \right),\\
\label{eq:9Bb}
&{P_2}\left( 1 \right) = \left( {{q_2}{a_1} + \left( {\begin{array}{*{20}{c}}
2\\
1
\end{array}} \right){q_2}\left( {1 - {q_2}} \right){a_2}{\rm{ + }}...{\rm{ + }}\left( {\begin{array}{*{20}{c}}
{{X_{\max }}}\\
1
\end{array}} \right){q_2}{{\left( {1 - {q_2}} \right)}^{{X_{\max }} - 1}}{a_{{X_{\max }}}}} \right){P_1}\left( 0 \right),\\
\nonumber & \vdots\\
\label{eq:9Be}
&{P_2}\left( {{X_{\max }}} \right) = q_2^{{X_{\max }}}{a_{{X_{\max }}}}{P_1}\left( 0 \right),
\end{align}
\end{subequations}
where $P_1(0) = 1$ as 0 molecules are received from the previous 0 time-slots.

\subsection{Problem Statement}
For the ISI channel, the average mutual information per channel use is given by
\begin{equation} \label{eq:3}
I = \mathop {\lim }\limits_{L \to \infty } \frac{1}{L}I\left( {{X^L};{Y^{L+k}}} \right),
\end{equation}
in which, ${X^L} = \left[ {{X_1},...,{X_L}} \right]$ and ${Y^{L+k}} = \left[ {{Y_1},...,{Y_{L+k}}} \right]$ , denote the length-$L$ input and $L+k$ output sequences, $k$ is the length of ISI, and the mutual information is evaluated for a given joint input distribution ${P_{{X^L}}}\left( {{x^L}} \right)$. This determines the achievable rate of reliable communication through this channel with this specific input distribution; the channel capacity is the supremum of this mutual information over all allowed joint input distributions.

Obtaining the capacity of the ISI channel requires optimization over the joint input distribution, a seemingly intractable problem. As suggested in~\cite{10}, we focus on the case of i.i.d. inputs, i.e., ${P_{{X^L}}} = {\bf{a}} \times ... \times {\bf{a}} = {{\bf{a}}^L}$, where ${\bf{a}} = \left[ {{a_0},...,{a_{{X_{\max }}}}} \right]$. We denote the resulting average mutual information as ${I_{i.i.d}}$~\cite{10}. Note that this mutual information is a lower bound on the expression in~\eqref{eq:3}, but, as in~\cite{10}, with a slight abuse of notation we will call the resulting mutual information the ``channel capacity". This capacity is given by
\begin{equation}\label{eq:4}
C = \mathop {\lim }\limits_{L \to \infty } \frac{1}{L}\mathop {\sup }\limits_{{{\bf{a}}^{L}}} I\left( {{X^L};{Y^{L+k}}} \right),
\end{equation}
Despite this significant simplification, calculating the channel capacity in \eqref{eq:4} appears intractable and, in the next sections, we investigate two related scenarios. We first ignore ISI and obtain the capacity of the resulting DMC under various possible constraints. In Section~\ref{sec:ISI}, we consider an ISI channel with a memory of one time-slot.

\section{Capacity Analysis of Molecular DMC with ASK} \label{sec:DMC}

We begin by analyzing several capacity measures with different cost constraints for the basic discrete memoryless molecular communication channel with ASK. As described in Section II, the DMC model may also be motivated by a molecular communication system where the life expectancy of molecules is equal to the time-slot duration. Specifically, we derive the capacity per channel use, per unit time and per unit cost with possible constraints on peak and average cost of transmission. In each case, the optimum input distribution and the resulting maximized capacity measure are also quantified.

\subsection{Capacity Per Channel Use} \label{subsec:CperChannelUse}
The general capacity problem with only a constraint on the maximum number of transmitted molecules per channel use (akin to a peak constraint on the input) is as follows
\begin{eqnarray}\label{eq:SC1}
        &\mathop {\sup }\limits_{P(X_m)} I\left( {X_m;Y_m} \right),\\
        \nonumber \mathrm{subject~to~~~}
        & 0 \le X_m \le {X_{\max }} \nonumber
\end{eqnarray}
where $X_{\max}$ is the allowed maximum number of transmitted molecules per time-slot. Note that for this case, we allow the transmitter to not release any molecules in a time-slot, i.e., $X=0$ is allowed. The objective function in~\eqref{eq:SC1} is concave with respect to the input probability vector~$\mathbf{a}$ and the constraint is linear, hence the optimization problem is concave. Hence, the solution of problem in~\eqref{eq:SC1} can be obtained using the Blahut-Arimoto algorithm~\cite{17}.

To describe the algorithm as applied here, using~\eqref{eq:2}, we define an $(X_{\max} + 1)\times (X_{\max}+1)$ channel transition matrix $\mathbf{P}$ as $\mathbf{P}_{x,y} = \left[ {p\left( {\left. {{Y_m} = {y}} \right|{X_m} = {x}} \right)} \right]$, where, ${y} \in \left[ {0,...,{X_{\max }}} \right]$ and ${x} \in \left[ {0,...,{X_{\max }}} \right]$; and matrix ${\bf{Q}} $ with $\mathbf{Q}_{y,x} = \left[ {p\left( {\left. {{X_m} = {x}} \right|{Y_m} = {y}} \right)} \right]$  with the same size as ${\bf{P}}$, where $\textbf{P}_{x,y}$ $\left(\mathbf{Q}_{y,x}\right)$ denotes the $(x,y)$$\left((y,x)\right)$-th element of $\textbf{P}$$\left(\mathbf{Q}\right)$. Let
\begin{equation}\label{eq:SC5}
J\left( {\textbf{a},{\bf{P}},{\bf{Q}}} \right) = \sum\nolimits_x {\sum\nolimits_y {{a_x}{\bf{P}}_{x,y}\log \frac{{{{\bf{Q}}_{y,x}}}}{{{a_x}}}} } .
\end{equation}
 Then the following is true~\cite{17}:
\begin{enumerate}
  \item $C = \mathop {\max }\limits_{\bf{Q}} \mathop {\max }\limits_{\bf{a}} J\left( {{\bf{a}},{\bf{Q}},{\bf{P}}} \right).$
  \item For fixed $\textbf{a}$, $J\left( {\textbf{a},{\bf{P}},{\bf{Q}}} \right)$ is maximized by
\begin{equation}\label{eq:SC60}
\mathbf{Q}_{y,x} = \frac{a_x \mathbf{P}_{x,y}}{\sum\nolimits_j a_x\mathbf{P}_{x,y}}.
\end{equation}
  \item For fixed ${\bf{Q}}$, $J\left( {\textbf{a},{\bf{P}},{\bf{Q}}} \right)$ is maximized by
      \begin{equation}\label{eq:SC6}
{a_x} = \frac{{\exp \left( \sum\nolimits_y {\mathbf{P}}_{x,y}\log {{\bf{Q}}_{y,x} } \right)}}{{\sum\nolimits_x {\exp \left( \sum\nolimits_y {\bf{P}}_{x,y}\log {{\bf{Q}}_{y,x} } \right) }}}, \hspace*{5ex} 0 \leq x \leq X_{\max}.
\end{equation}
\end{enumerate}
The Blahut-Arimoto algorithm begins with the transition probability matrix defined by~\eqref{eq:2} and an arbitrary, but valid, choice for $\mathbf{a}$. It then iterates between steps 2 and 3 above until convergence. Since, the mutual information in~\eqref{eq:SC1} is concave in terms of input probability, the output of the algorithm is the optimal, capacity-achieving, input probability distribution, $\hat{\mathbf{a}}$; as an aside, the matrix $\mathbf{Q}$ is the corresponding output-input reverse transition matrix.

\begin{figure}
        \centering
        \vspace{-1.5\baselineskip}
        \begin{subfigure}[b]{0.49\textwidth}
                \caption{}
                \includegraphics[width=\textwidth]{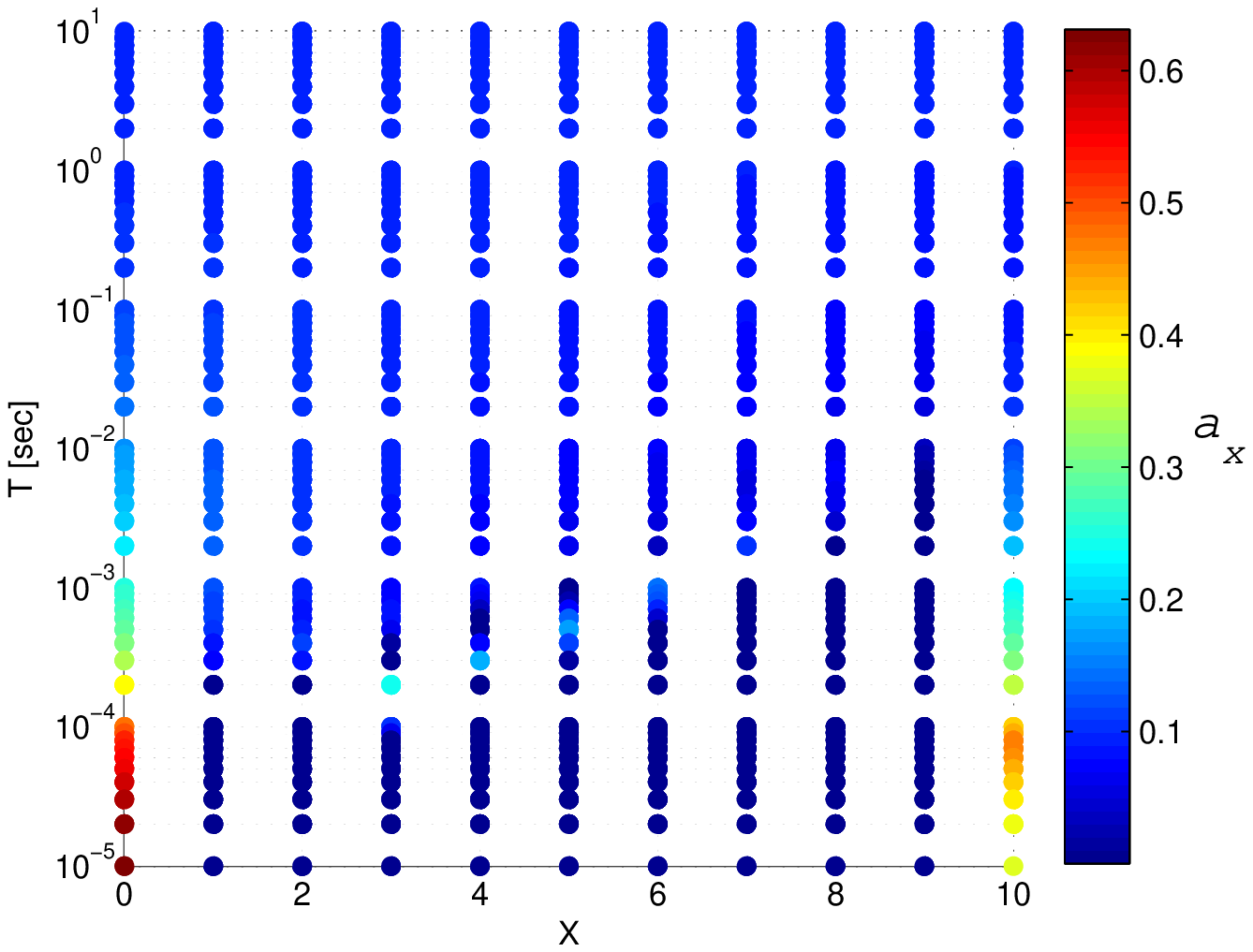}
                \label{fig:input_dis}
        \end{subfigure}
        \begin{subfigure}[b]{0.49\textwidth}
                \caption{}
                \includegraphics[width=\textwidth]{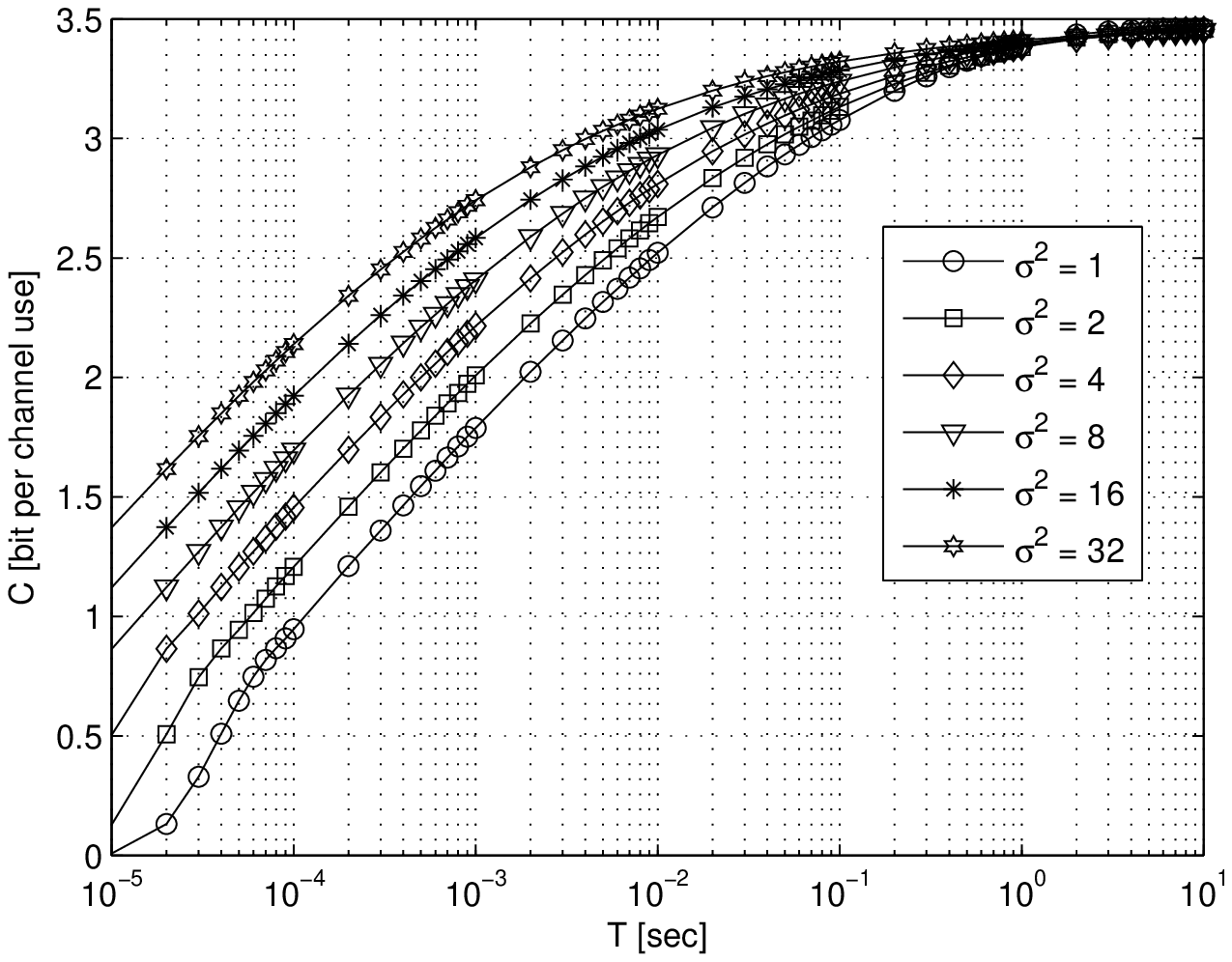}
                \label{fig:C_sigma}
        \end{subfigure}\\%
        \vspace{-1.5\baselineskip}
        \begin{subfigure}[b]{0.49\textwidth}
                \caption{}
                \includegraphics[width=\textwidth]{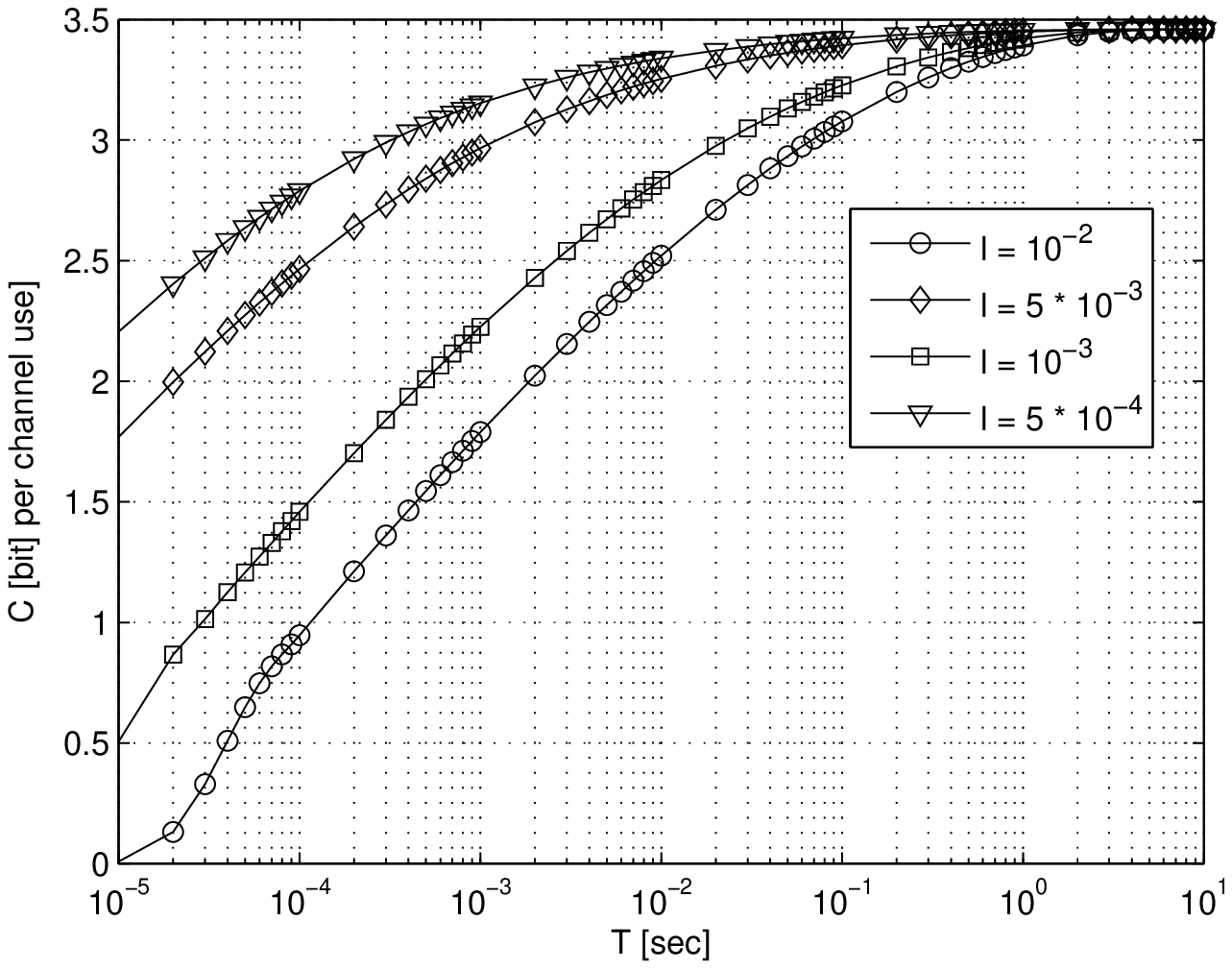}
                \label{fig:Cl}
        \end{subfigure}
        \begin{subfigure}[b]{0.49\textwidth}
                \caption{}
                \includegraphics[width=\textwidth]{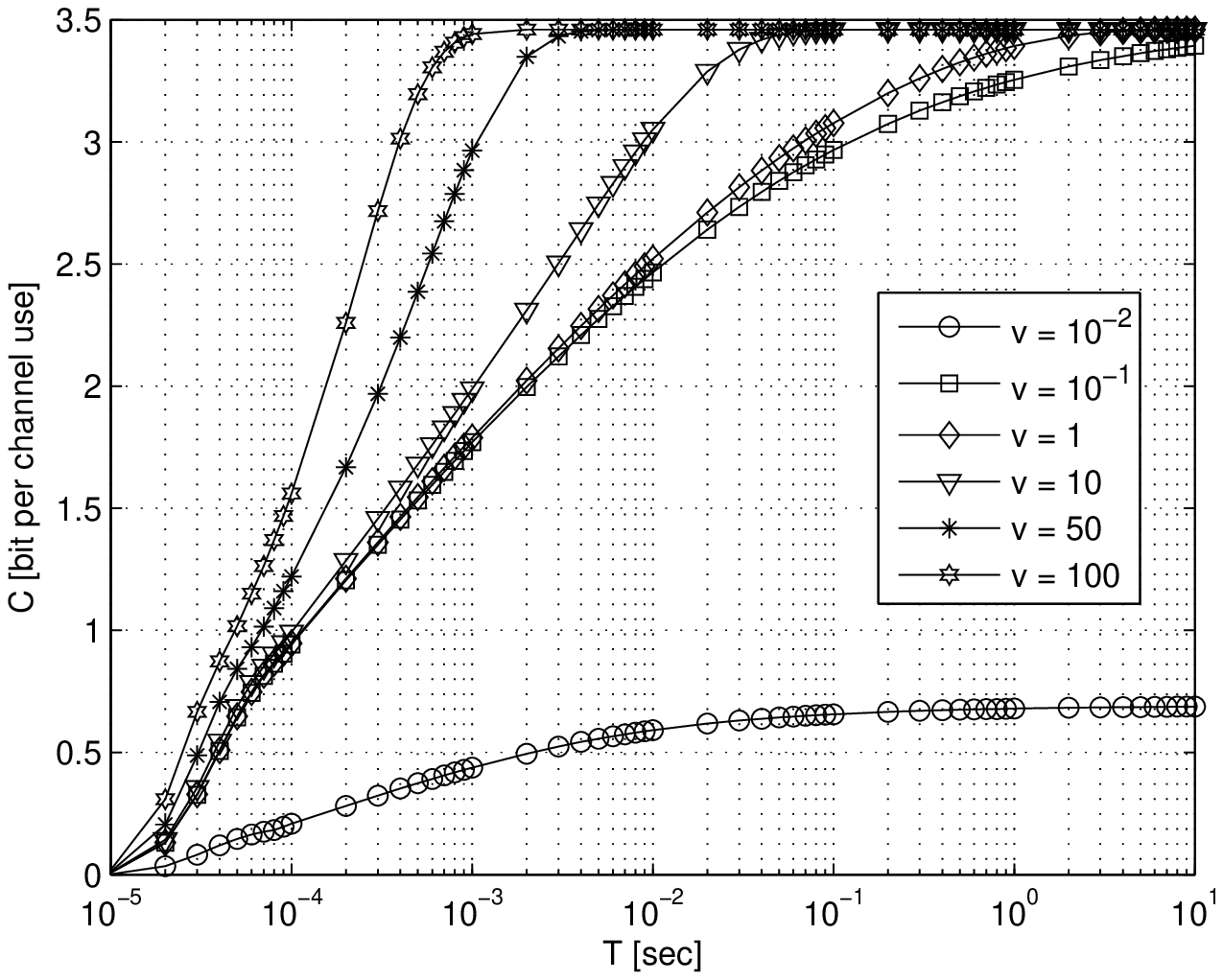}
                \label{fig:C_v}
        \end{subfigure}
        \vspace{-1.5\baselineskip}
%        \begin{subfigure}[b]{0.33\textwidth}
%                \caption{}
%                \includegraphics[width=\textwidth]{fig_20a.eps}
%                \label{fig:C_Xmax}
%        \end{subfigure}%
%        \begin{subfigure}[b]{0.33\textwidth}
%                \caption{}
%                \includegraphics[width=\textwidth]{fig_20b.eps}
%                \label{fig:CR_Xmax}
%        \end{subfigure}
        ~ %add desired spacing between images, e. g. ~, \quad, \qquad, \hfill etc.
        \caption{(a) Optimum $a_x$ in terms of $T$ with $l=10^{-2}$, $v = 1$, $\sigma^2 = 1$. (b-d) $C$ in terms of $T$ for different values of (b) $\sigma^2$ with $v = 1$ and $l = 10^{-2}$, (c) $l$ with $\sigma^2 = 1$ and $v = 1$ and (d) $v$ with $l = 10^{-2}$ and $\sigma^2 = 1$, all with $X_{\max}=10$. }
        \label{fig:Cap_ch_use}
\end{figure}

Figure~\ref{fig:Cap_ch_use} illustrates the results of using the Blahut-Arimoto algorithm in obtaining the optimal input distribution. Figure~\ref{fig:input_dis} plots the optimum input distribution, $a_x$, in terms of $T$ with $l=10^{-2}$, $v=1$, $\sigma^2=1$ and $X_{max}=10$. Interestingly, though perhaps intuitively, the optimal distribution transitions from a bipolar distribution for small $T$ to a uniform distribution for large $T$ (system essentially perfect). In all cases in the figure, as $T$ increases, the capacity converges to $\log_2(X_{\max} + 1)$.

Figure~\ref{fig:C_sigma} plots the corresponding capacities as a function of the time-slot duration, $T$, for different values of $\sigma^2$ and the same, fixed, values of $l$ and $v$. We observe that, for the given values of $v$ and $l$, by increasing $\sigma^2$ and $T$, the capacity increases. While it is intuitive that increasing $T$ should increase the capacity per channel use, by increasing the probability of receiving all molecules in one time-slot, the increase with increasing $\sigma^2$ is less intuitive. The explanation is that increasing the diffusion constant increases the randomness in the location of the molecules, and the randomness helps improve the chances of molecules arriving at the destination within one time-slot. This is consistent with the fact that $q_1 = F_W(T)$ is an increasing function of $\sigma$ for low-to-medium values of $v$. For large $v$, $q_1 \simeq 1$ and capacity is near its maximum value. It is worth noting that when information is encoded in time-of-release as in~\cite{2}, the mutual information is not monotonic in $\sigma$. %(later, in Fig.~\ref{fig:cap_time}, we will illustrate a similar behavior in our case).

Figures~\ref{fig:Cl} and~\ref{fig:C_v} plot the capacity for different values of $l$ and $v$, respectively. As expected, decreasing $l$ (equivalently, increasing $v$) improves capacity significantly because, in each case, clearly, $q_1$ increases.

\subsection{Capacity with Average Cost Constraint} \label{subsec:CwithCostConstraints}

A variation on the optimization in~\eqref{eq:SC1} is when the average cost of transmitted molecules per channel use is also constrained. The resulting optimization problem is
\begin{subequations}\label{eq:SC7}
      \begin{align}
        \label{eq:SC7a}
        &\mathop {\sup }\limits_{p(X_m)} I\left( {X_m;Y_m} \right),\\
        \label{eq:SC7b}
        \mathrm{subject~to~~~} & 0 \le X_m \le {X_{\max }}\\
        \label{eq:SC7c}
        & \hspace*{-0.3in} \texttt{E}\left( e_{X_m} \right) \le E
        \end{align}
\end{subequations}
where $e_{X_m=x}\ge0$ is the cost of using $x$ molecules as the ASK symbol. The vector ${\bf{e}} = \left[ {{e_x}} \right]$, $x \in \left\{ {0,1,...,{X_{\max }}} \right\}$ specifies the cost vector. A concrete example is when the transmission cost is equal (proportional) to the number of transmitted molecules per channel use, i.e., ${e_x} = x$. The capacity with cost constraint $E$ is then defined as
\begin{equation}\label{eq:SC8}
C\left( E \right) = \mathop {\max }\limits_{\textbf{a} \in {A_E}} \sum\nolimits_x {\sum\nolimits_y {{a_x}{{\bf{P}}_{x,y}}\log \frac{{{{\bf{P}}_{x,y}}}}{{\sum\nolimits_x {{a_x}{{\bf{P}}_{x,y}}} }}} }  = \mathop {\max }\limits_{\mathbf{a} \in {A_E}} I\left( {\textbf{a},{\bf{P}}} \right)
\end{equation}
where $A_E = \left\{ \mathbf{a} \mid \sum\nolimits_x {{a_x}{e_x}}  \le E \right\}$ denotes the set of all allowable input distributions that meet the cost constraint.

With a concave cost function and linear constraints, the optimization problem in~\eqref{eq:SC8} is concave. Hence, the solution can be obtained using the constrained Blahut-Arimoto algorithm~\cite{17}. Using Lagrange multipliers, the cost function can be parameterized as
\begin{equation}\label{eq:SC10}
C\left( E \right) = \mathop {\max }\limits_\textbf{a} \left[ {\sum\nolimits_x {\sum\nolimits_y {{a_x}{{\bf{P}}_{x,y}}\log \frac{{{{\bf{P}}_{x,y}}}}{{\sum\nolimits_x {{a_x}{{\bf{P}}_{x,y}}} }}} }  - s\left( {\sum\nolimits_x {{a_x}{e_x} - E} } \right)} \right],
\end{equation}
where $s$ denotes the Lagrange multiplier. The maximization is now over all input probability vectors $\textbf{a}$. For any matrix ${\bf{Q}}$ with size $\left( {{X_{\max }} + 1} \right) \times \left( {{X_{\max }} + 1} \right)$, let
\begin{equation}\label{eq:SC12}
J\left(s, {\textbf{a},{\bf{P}},{\bf{Q}}} \right) = \sum\nolimits_x {\sum\nolimits_y {{a_x}{{\bf{P}}_{x,y}}\log \frac{{{{\bf{Q}}_{y,x}}}}{{{a_x}}} - s\sum\nolimits_x {{a_x}{e_x}} } }
\end{equation}
Hence, using \eqref{eq:SC12}, the following is true~\cite{17}
\begin{enumerate}
  \item The constrained capacity $C= s\sum_x x a_x + \mathop {\max }\limits_{\bf{Q}} \mathop {\max }\limits_\textbf{a} J\left( s, {\mathbf{a},{\bf{P}},{\bf{Q}}} \right)$
  \item For fixed $\textbf{a}$, $J\left(s, {\textbf{a},{\bf{P}},{\bf{Q}}} \right)$ is maximized by
      \begin{equation}\label{eq:SC13}
            {{\bf{Q}}_{y,x}} = \frac{{{a_x}{{\bf{P}}_{x,y}}}}{{\sum\nolimits_x {{a_x}{{\bf{P}}_{x,y}}} }}
      \end{equation}
  \item For fixed $s$, $\textbf{P}$, $J\left( s, {\textbf{a},{\bf{P}},{\bf{Q}}} \right)$ is maximized by
      \begin{equation}\label{eq:SC14}
            {a_x} = \frac{{\exp \left( {\sum\nolimits_y {{{\bf{P}}_{x,y}}\log {{\bf{Q}}_{y,x}} - s{e_x}} } \right)}}{{\sum\nolimits_x {\exp \left( \sum\nolimits_y {{{\bf{P}}_{x,y}}\log {{\bf{Q}}_{y,x} - s{e_x}} } \right) }}}.
      \end{equation}
  \item The optimum $s$ is the minimum value ($\geq 0$) that satisfies the constraint $\sum_x e_x a_x \leq E$.
  \end{enumerate}

The procedure to optimize for the input distribution is similar to that without the average cost constraint. The one additional step is to obtain the parameter $s$ after updating the distribution vector $\mathbf{a}$. As per 4) above, the minimum value of $s$ can be obtained by a bisection search. To obtain the optimal distribution, the algorithm is executed setting $E = \sum_x e_x a_x $. In the rest of the paper, unless otherwise mentioned, we consider $e_x=x$.

\begin{figure}
        \centering
        \vspace{-1.5\baselineskip}
        \begin{subfigure}[b]{0.49\textwidth}
                \caption{}
                \includegraphics[width=\textwidth]{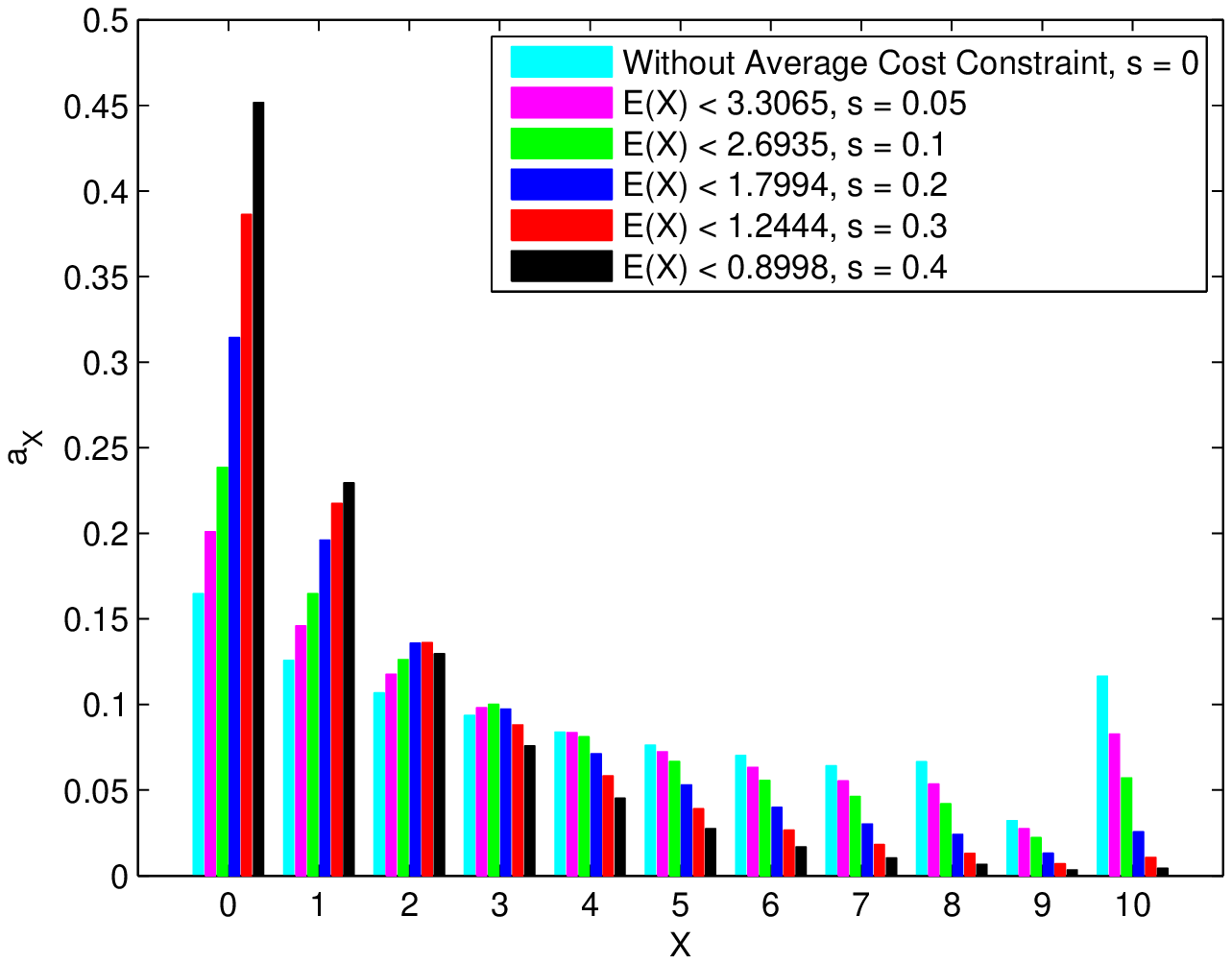}
                \label{fig:opt_in10}
        \end{subfigure}
        \begin{subfigure}[b]{0.49\textwidth}
                \caption{}
                \includegraphics[width=\textwidth]{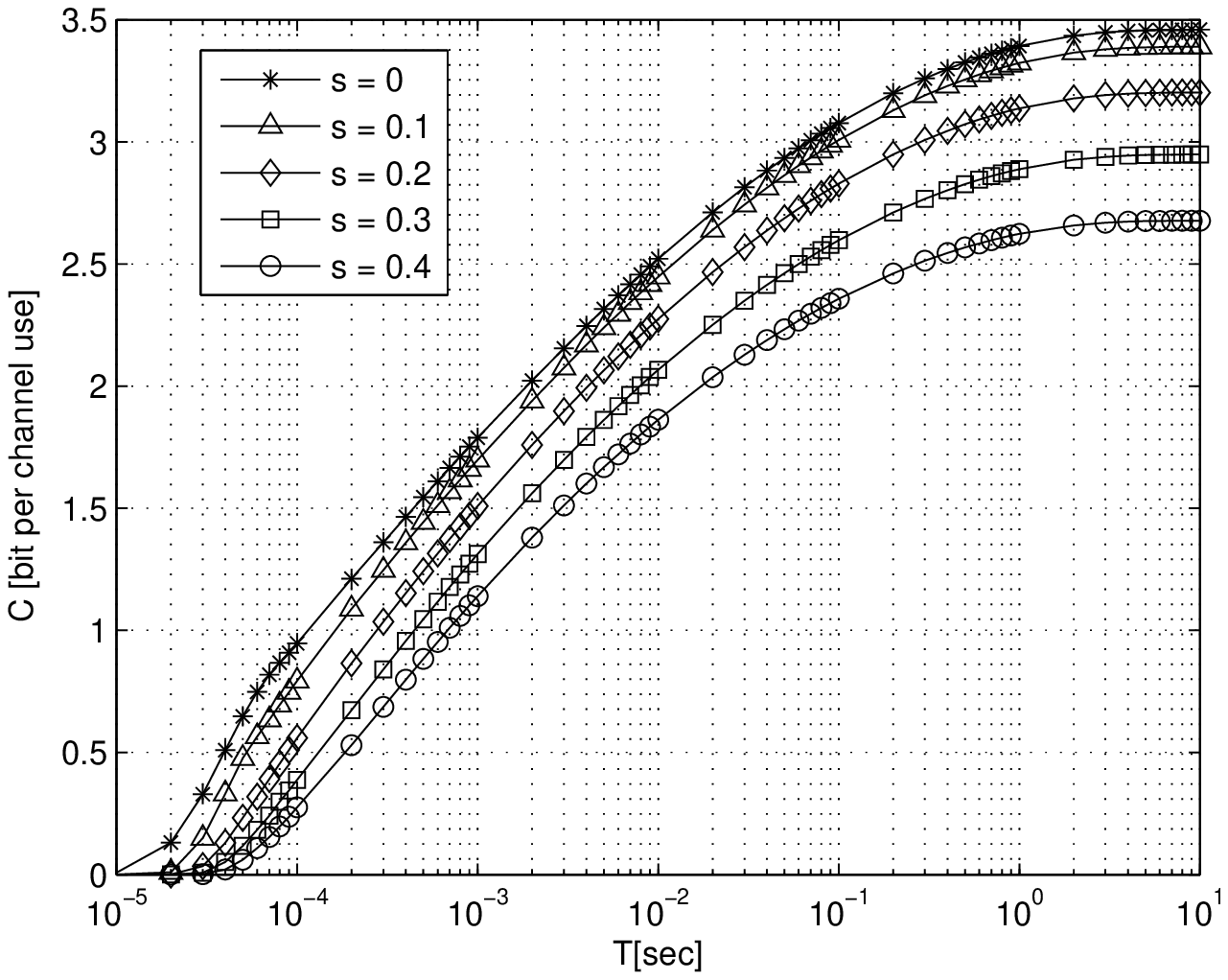}
                \label{fig:C_ave}
        \end{subfigure}%
        \vspace{-1.5\baselineskip}
        ~ %add desired spacing between images, e. g. ~, \quad, \qquad, \hfill etc.
        \caption{(a) Optimum $a_x$ for different values of average cost constraint with $T = 10 msec$, (b) $C$ in terms of average cost constraint for different values of $T$, all with $\sigma^2 = 1$, $v = 1$, $l = 10^{-2}$, $X_{\max} = 10$.}
        \label{fig:Cap_ch_use_av}
\end{figure}

Figure~\ref{fig:Cap_ch_use_av} plots the result of including an average cost constraint within the optimization problem of ~\eqref{eq:SC7}. For ease of illustration, in Fig.~\ref{fig:opt_in10}, with $T = 10$~msec, we vary the Lagrange multiplier $s$, to vary the average transmission cost. Increasing the parameter $s$ in \eqref{eq:SC14} reduces $E = \texttt{E}(e_{x})$ as expected; further, reducing $E$, i.e., making transmissions expensive, shifts the optimal probability distribution to the lower values of $x$. Figure.~\ref{fig:C_ave} plots the constrained capacity in bits per channel use in terms of average cost constraint, i.e. $E$, in molecules, for different values of $T$. It is evident that by increasing $E$ and $T$ the capacity increases, i.e., as expected, a stricter constraint reduces the achievable rate.

\subsection{Capacity Per Unit Time} \label{subsec:CperUnitTime}

It is worth noting that the solutions to the optimization problems in Sections~\ref{subsec:CperChannelUse} and~\ref{subsec:CwithCostConstraints} can be used to solve other problems of interest such as to maximize the mutual information per unit time with constraints on the maximum and/or average number of transmitted molecules per unit time. The general optimization problem is as follows
\begin{subequations}\label{eq:SC16}
\begin{align}\label{eq:SC16a}
        &\mathop {\sup }\limits_{p(X_m)} \frac{{I\left( {{X_m};{Y_m}} \right)}}{T}\\\label{eq:SC16b}
\mathrm{subject~to~~~} &0 \le \frac{{{X_m}}}{T} \le {P_{\max }},\\\label{eq:SC16c}
        &0 \le E\left( {\frac{{{e_{X_m}}}}{T}} \right) \le \bar P
\end{align}
\end{subequations}
where, $P_{\max }$ and $\bar P$ are the maximum and the average number of transmitted molecules allowed \emph{per unit time}. However, since $T$ is a constant and independent of $a_x$, we can use the same approach as described above by setting $X_{\max}$ to be the integer closest to $\left(P_{\max}T\right)$ and $E = \bar{P}T$.

% Figure 3
 \begin{figure}
        \centering
        \vspace{-1.5\baselineskip}
        \begin{subfigure}[b]{0.49\textwidth}
                \caption{}
                \includegraphics[width=\textwidth]{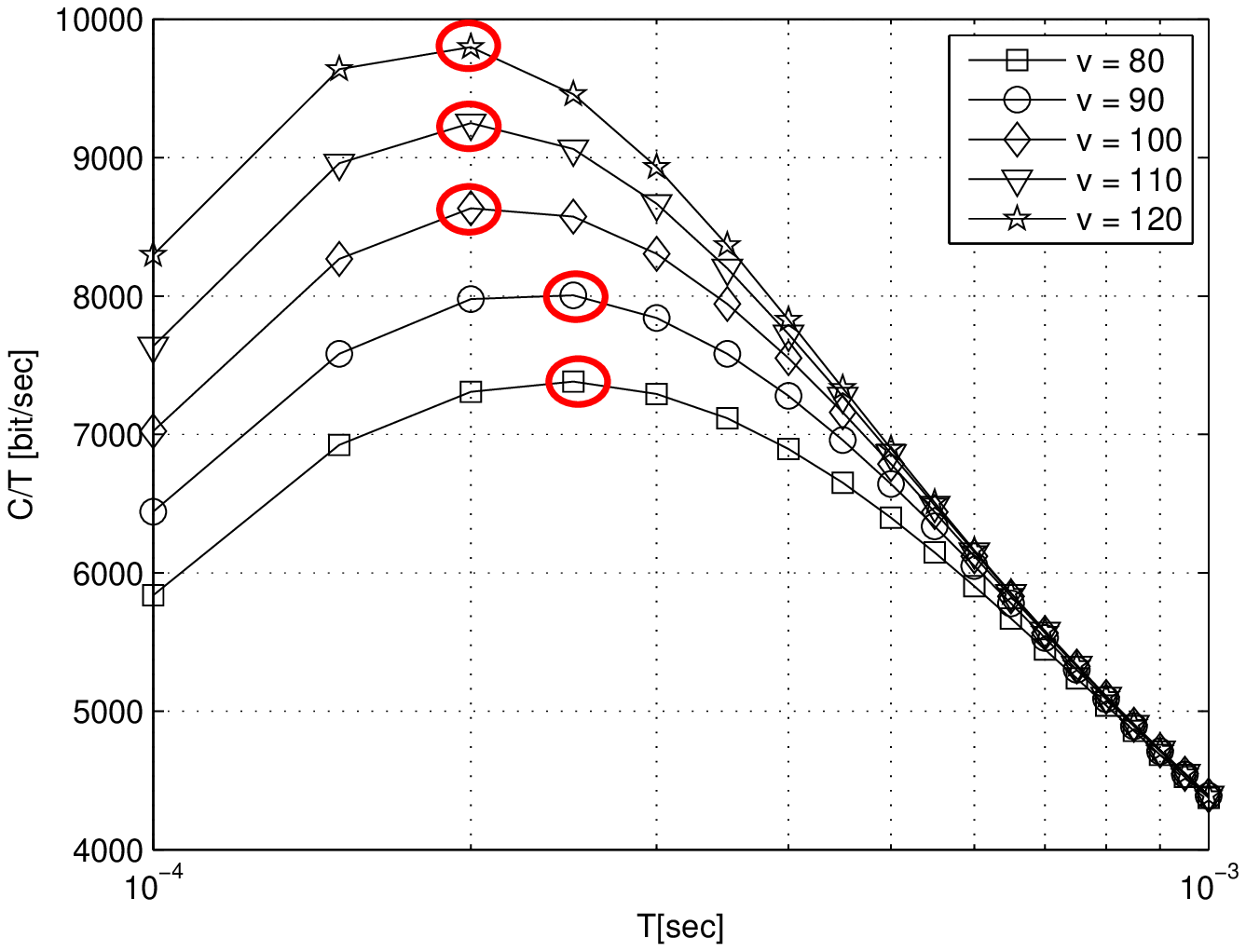}
                \label{fig:cap_v}
        \end{subfigure}%
        ~ %add desired spacing between images, e. g. ~, \quad, \qquad, \hfill etc.
        \begin{subfigure}[b]{0.49\textwidth}
                \caption{}
                \includegraphics[width=\textwidth]{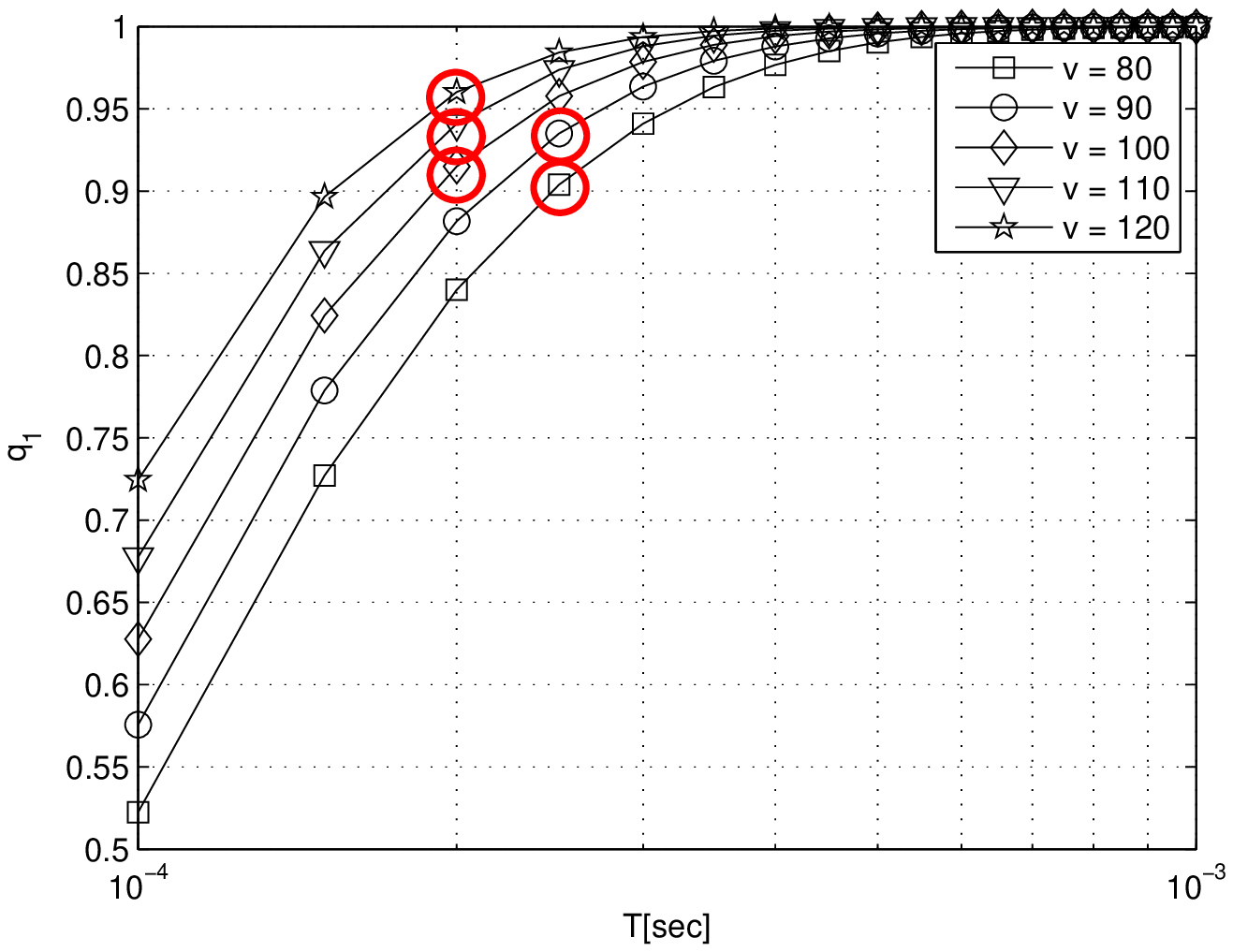}
                \label{fig:q_v}
        \end{subfigure}\\
        \vspace{-1.5\baselineskip}
        \begin{subfigure}[b]{0.49\textwidth}
                \caption{}
                \includegraphics[width=\textwidth]{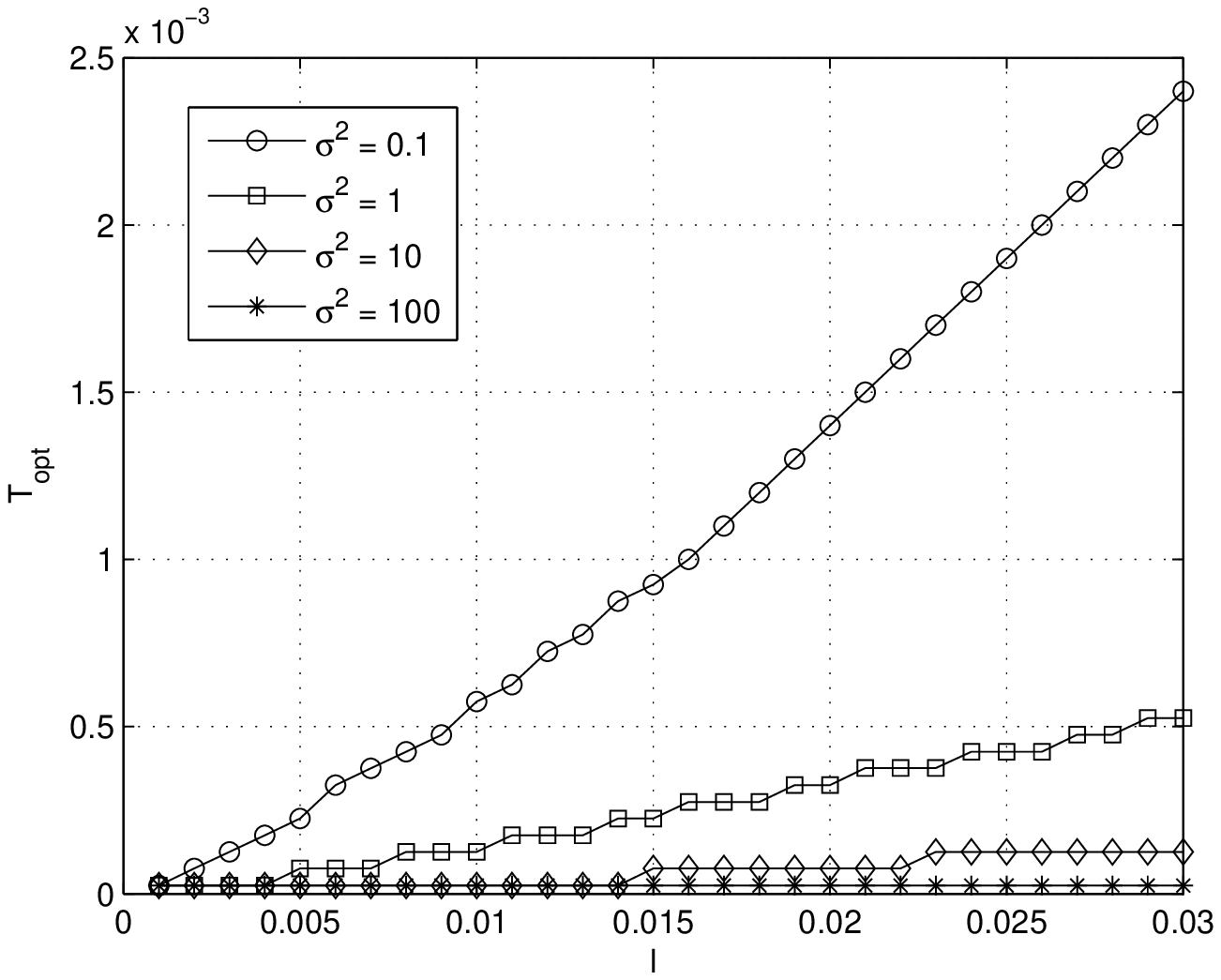}
                \label{fig:Topt}
        \end{subfigure}
 %       \vspace{-1.5\baselineskip}
%        \begin{subfigure}[b]{0.49\textwidth}
%                \caption{}
%                \includegraphics[width=\textwidth]{fig_3a.eps}
%                \label{fig:cap_sig}
%        \end{subfigure}%
        ~ %add desired spacing between images, e. g. ~, \quad, \qquad, \hfill etc.
%        \begin{subfigure}[b]{0.49\textwidth}
%                \caption{}
%                \includegraphics[width=\textwidth]{fig_3b.eps}
%                \label{fig:q_sig}
%        \end{subfigure}
        \vspace{-1.5\baselineskip}
        \caption{(a) $C/T$ and (b) $q_1$ in terms of $T$ for different values of $v$ with $\sigma^2 = 0.5$, $l = 10^{-2}$, (c) $T_{opt}$ in terms of $l$ for different values of $\sigma^2$ with $v$ = 1}\label{fig:cap_time} %(c) $C/T$ and (d) $q_1$ in terms of $T$ for different values of $\sigma^2$ with $\l = 10^{-2}$, $v = 100$ all with $X/T  \le 20 \times {10^3}$ molecules/sec, Red markers show the maximum value of $C/T$ and its corresponding value of $q$.}
\end{figure}

We note that a related problem of interest is obtaining the optimal time-slot duration, $T_{opt}$, that maximizes the capacity per unit time for a given set of system parameters ($v$, $l$ and $\sigma^2$). Unfortunately, this optimization problem is not concave and so we have resorted here to numerical solutions.

% \begin{figure}
%        \centering
%        \vspace{-1.5\baselineskip}
%        \begin{subfigure}[t!]{0.49\textwidth}
%                \includegraphics[width=\textwidth]{Figure_T_opt.eps}
%
%        \end{subfigure}%
%      %  \vspace{-0.5\baselineskip}
%        \caption{$T_{opt}$ in terms of $l$ for different values of $\sigma^2$}\label{fig:Topt}
%        \vspace{-1.5\baselineskip}
% \end{figure}

Figures~\ref{fig:cap_v} and~\ref{fig:q_v} plot $C/T$ in bits/sec and $q_1$, respectively, as functions of time-slot duration $T$, for different values of $v$. Here, $\sigma^2 = 0.5$, $l = 10^{-2}$ and $X/T  \le 20 \times {10^3}$ molecules/sec. As expected, there is an optimum value for time-slot duration, $T_{opt}$, which maximizes the $I(X_m;Y_m)/T$, though, interestingly, this duration does not vary significantly with drift velocity. As drift velocity increases, the optimal value slowly decreases. Locating the corresponding values of $q_1$ in Fig~\ref{fig:q_v} we observe that the maximum value of $C/T$ corresponds to a value of $q_1$ higher than 0.9. For such a large probability that the molecules arrive within the same time-slot, the assumed DMC model turns out to be essentially accurate. Hence, we plot $C/T$ in Figure~\ref{fig:cap_v} for the range of $T$ which corresponds to this value of $q_1$.

Figure~\ref{fig:Topt} plots $T_{opt}$, derived numerically, as a function of $l$ for different values of $\sigma^2$. We observe that increasing $\sigma^2$, $T_{opt}$ reduces while it increases with increasing $l$ (to ensure that an adequate percentage of molecules arrive within the time-slot).

%Figure 4
\begin{figure}
        \centering
        \vspace{-1.5\baselineskip}
        \begin{subfigure}[t!]{0.49\textwidth}
                \caption{}
                \includegraphics[width=\textwidth]{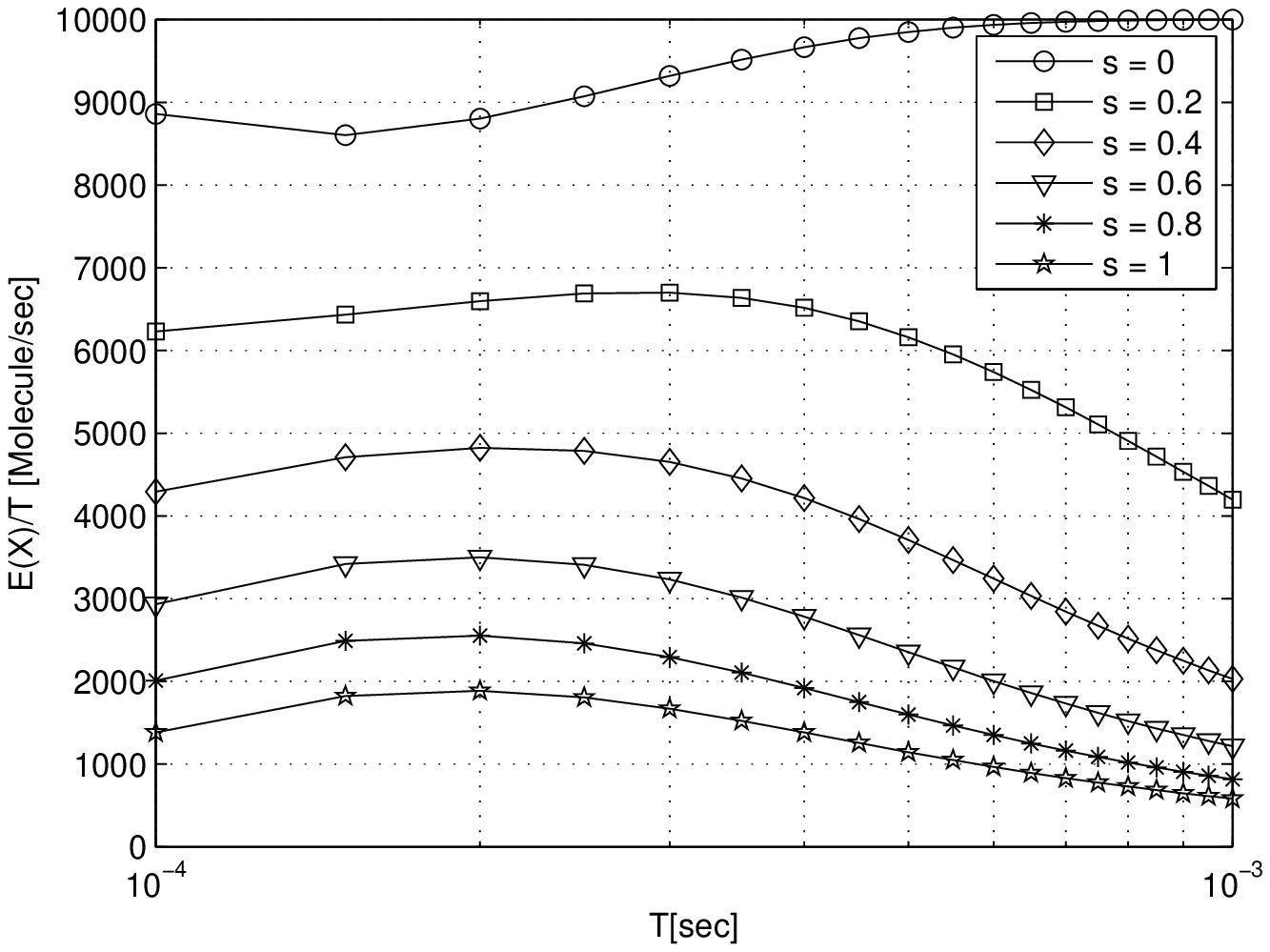}
                \label{fig:cost_cost1}
        \end{subfigure}%
        ~ %add desired spacing between images, e. g. ~, \quad, \qquad, \hfill etc.
        \begin{subfigure}[t!]{0.49\textwidth}
                \caption{}
                \includegraphics[width=\textwidth]{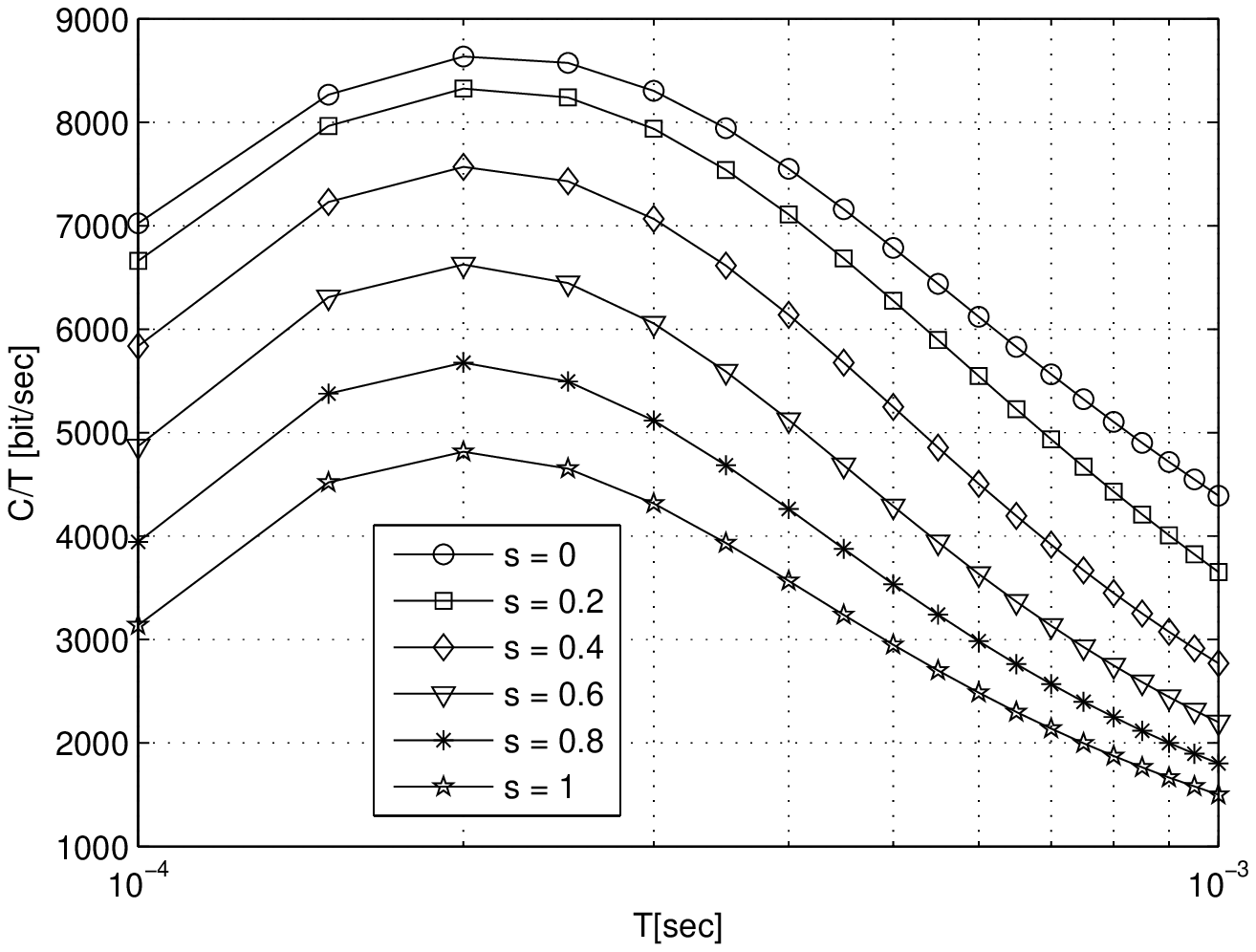}
                \label{fig:cap_cost_1}
        \end{subfigure}
        \\
        \vspace{-1.5\baselineskip}
        \caption{(a) $E(X/T)$ (b) $C/T$ in terms of $T$, for different values of $s$ and $e(X) = X$, all with constraint on average transmuted molecules per second and $\l = 10^{-2}$, $v = 100$ $\sigma^2 =0.5$ and $X/T  \le 20 \times {10^3}$ molecules/sec.}\label{fig:cap_p_time_ave}
        \label{fig:ave_con}
\end{figure}

Figures~\ref{fig:cost_cost1} and \ref{fig:cap_cost_1} plot $E(X/T)$ and $C/T$ in terms of $T$ for different values of $s$. In the results of these figures, the input distribution is derived from optimization problem in~\eqref{eq:SC16}. Here $E(X/T)$ and $C/T$ decrease with increasing $s$. Since, $s$ is a Lagrange multiplier, in~\eqref{eq:SC7}, increasing its value tightens the constraint in~\eqref{eq:SC7c}, i.e., $E(X/T)$ and $C/T$ are also reduced.

\subsection{Capacity Per Unit Cost}

A final optimization problem within the DMC framework of interest is the relative capacity, i.e., the capacity per unit cost. This is akin to optimizing the ``energy efficiency" in molecular communications. Here, we investigate maximizing the mutual information per average transmission cost when the maximum number of transmitted molecules per channel use is limited. The associated optimization problem is given by
\begin{eqnarray}\label{eq:SC19}
        &\mathop {\sup }\limits_{p(X_m)} \frac{{I\left( {{X_m};{Y_m}} \right)}}{\texttt{E}(e_{X_m})}         \label{eq:SC19A}
    \\
         \mathrm{subject~to~~~} & 0 \le X_m \le {X_{\max }} \nonumber
\end{eqnarray}
The ratio of mutual information to transmission cost, as in~\eqref{eq:SC19}, is defined in~\cite{18} as relative capacity, usually denoted by ${C_R}$.

In~\eqref{eq:SC19} the choice of cost $e_{X_m=x} = x$ is the average number of molecules transmitted, $E(X)$; allowing $X = 0$ implies that we allow for a symbol with zero cost. Clearly, any optimization would maximize use of the symbol ($a_0 = 1$). To avoid form zero information rate, we can add a constant, i.e., use $e_x = c_0 + x$, where $c_0$ is a small constant value, ensuring that there is no symbol with zero cost.

Since the relative capacity, $I(X_m;Y_m)/E(e_{X_m})$, is a continuous and quasi concave function of $\textbf{a}$, it may be maximized using the Jimbo-Kunisawa algorithm~\cite{19}. The corresponding iterative procedure is as follows:
\begin{enumerate}
   \item Initially, choose an arbitrary probability vector $\textbf{a}^{(0)}$,
  \item After the $r$th iteration, having obtained probability vector ${{\bf{a}}^{\left( r \right)}}$, construct the  $\left( {r + 1} \right)$th probability vector ${{\bf{a}}^{\left( r + 1 \right)}}$  as follows;
\begin{equation}\label{eq:SC20}
\tilde a_x^{\left( {r + 1} \right)} = a_x^{\left( r \right)}\exp \left[ {\frac{{e_0 {D_x}\left( {{{\bf{a}}^{\left( r \right)}}} \right)}}{{{e_x}}}} \right],{\rm{      }}x = 0,...,X_{\max},
\end{equation}
where
\begin{equation}\label{eq:SC21}
{D_x}\left( {{{\bf{a}}^{\left( r \right)}}} \right) = \sum\nolimits_y {{{\bf{P}}_{x,y}}} \log \frac{{{{\bf{P}}_{x,y}}}}{{\sum\nolimits_x {a_x^{\left( r \right)}{{\bf{P}}_{x,y}}} }}.
\end{equation}
Finally, we need to normalize the distribution using
\begin{equation}\label{eq:SC23}
a_x^{\left( {r + 1} \right)} = \frac{{\tilde a_x^{\left( {r + 1} \right)}}}{{\sum\nolimits_x {\tilde a_x^{r + 1}} }}.
 \end{equation}
\end{enumerate}
We iterate until $\left|C_R^{(r + 1)} - C_R^{(r)}\right|<\varepsilon$, where $C_R^{(r)}$ is the value of objective function in \eqref{eq:SC19A} at $r$th iteration and $\varepsilon>0$ is arbitrary small number.
%Figure 5
\begin{figure}
        \centering
        \vspace{-1.5\baselineskip}
        \begin{subfigure}[b]{0.49\textwidth}
                \caption{}
                \includegraphics[width=\textwidth]{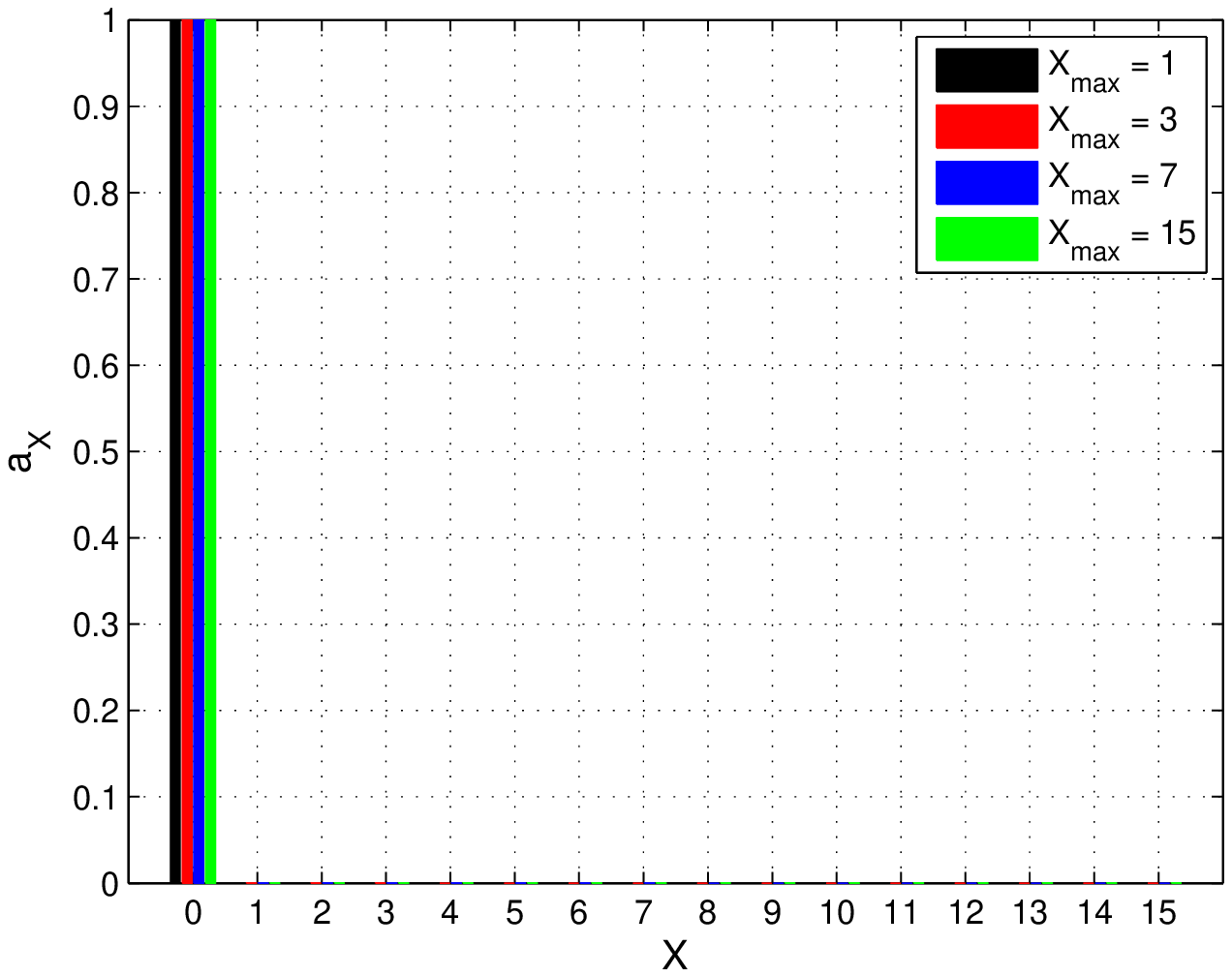}
                \label{fig:inp_rel_cap}
        \end{subfigure}
        \begin{subfigure}[b]{0.49\textwidth}
                \caption{}
                \includegraphics[width=\textwidth]{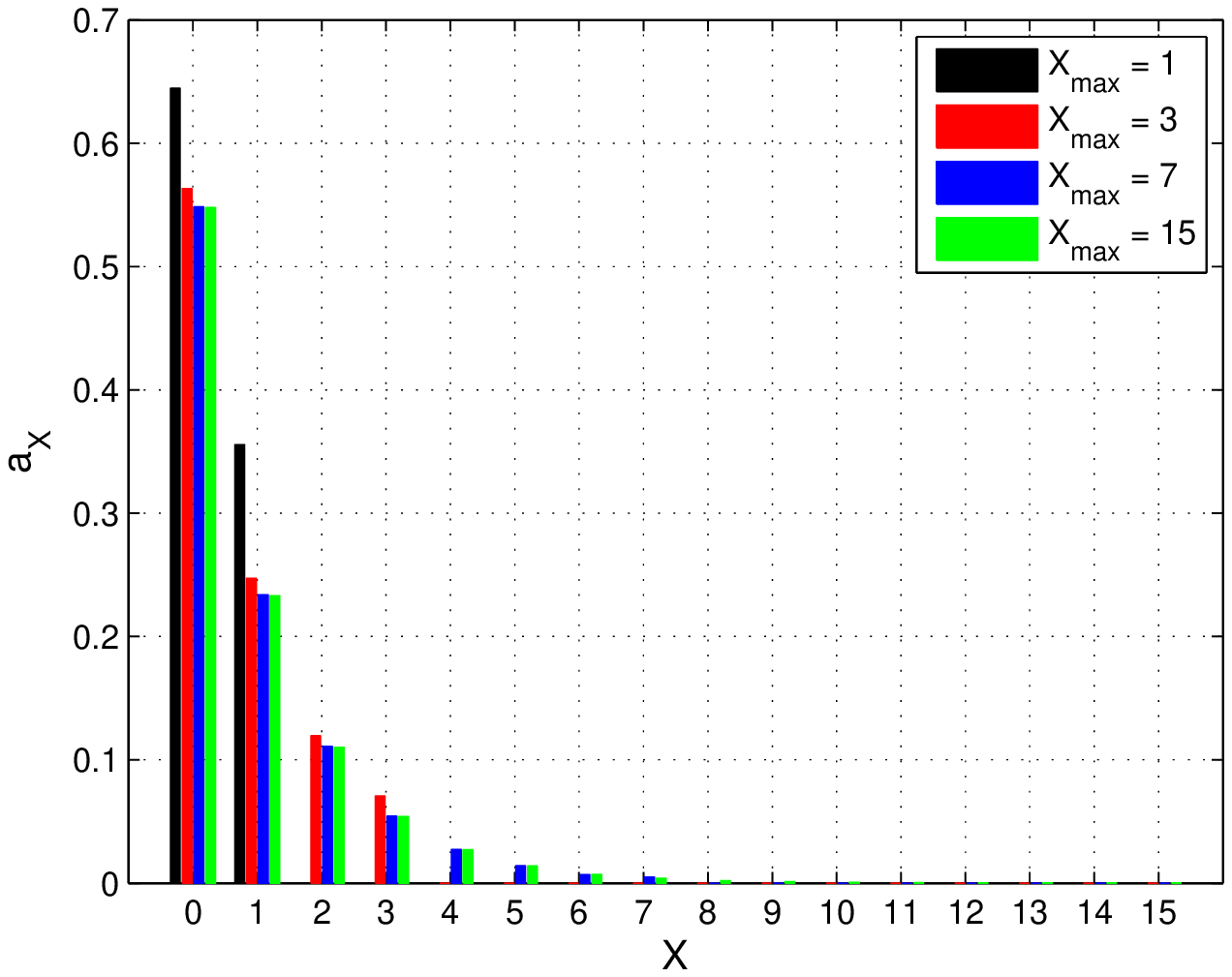}
                \label{fig:IN_cos10}
        \end{subfigure}\\%
        \vspace{-1.5\baselineskip}
        \begin{subfigure}[b]{0.49\textwidth}
                \caption{}
                \includegraphics[width=\textwidth]{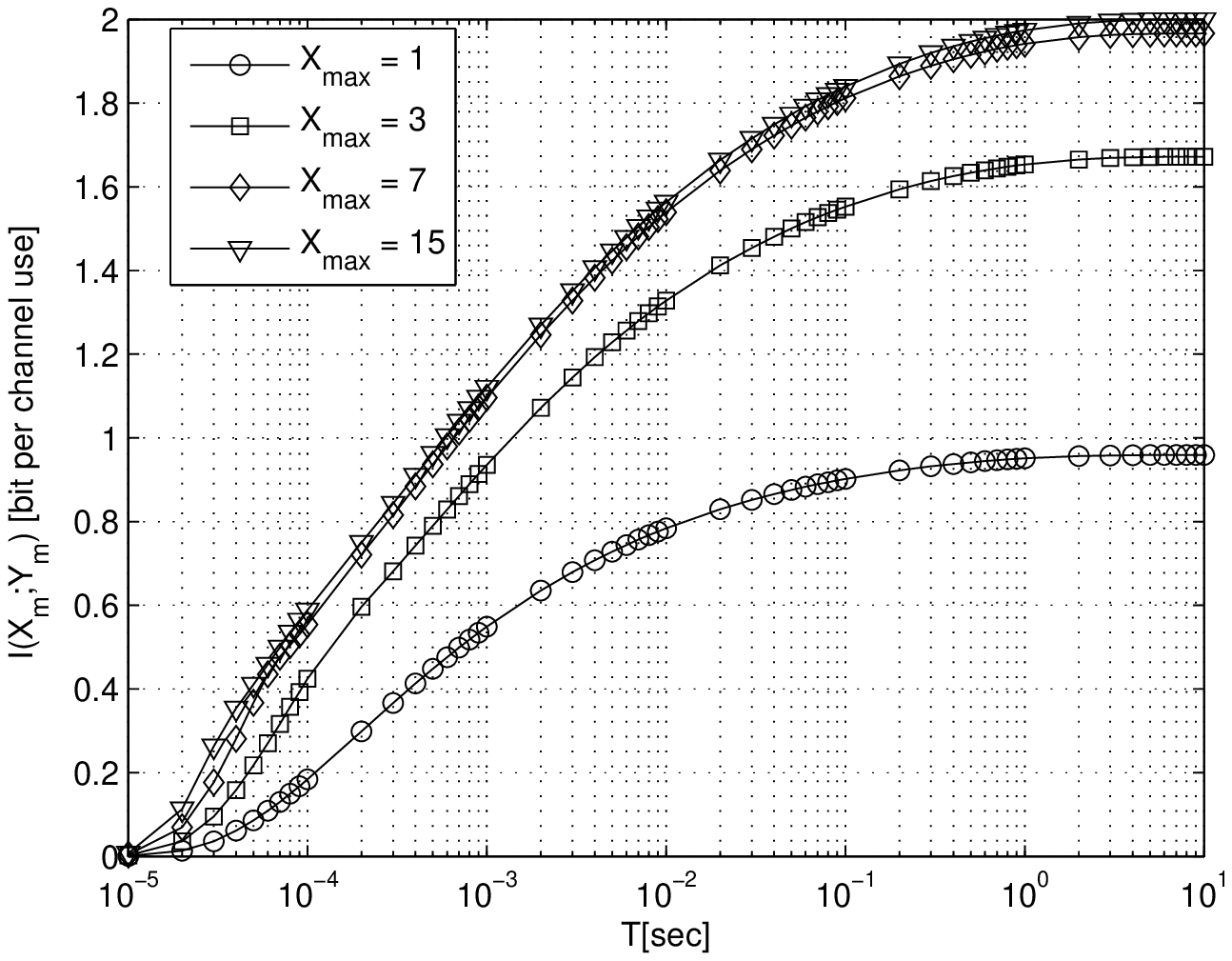}
                \label{fig:IXY2}
        \end{subfigure}%
        ~ %add desired spacing between images, e. g. ~, \quad, \qquad, \hfill etc.
        \begin{subfigure}[b]{0.49\textwidth}
                \caption{}
                \includegraphics[width=\textwidth]{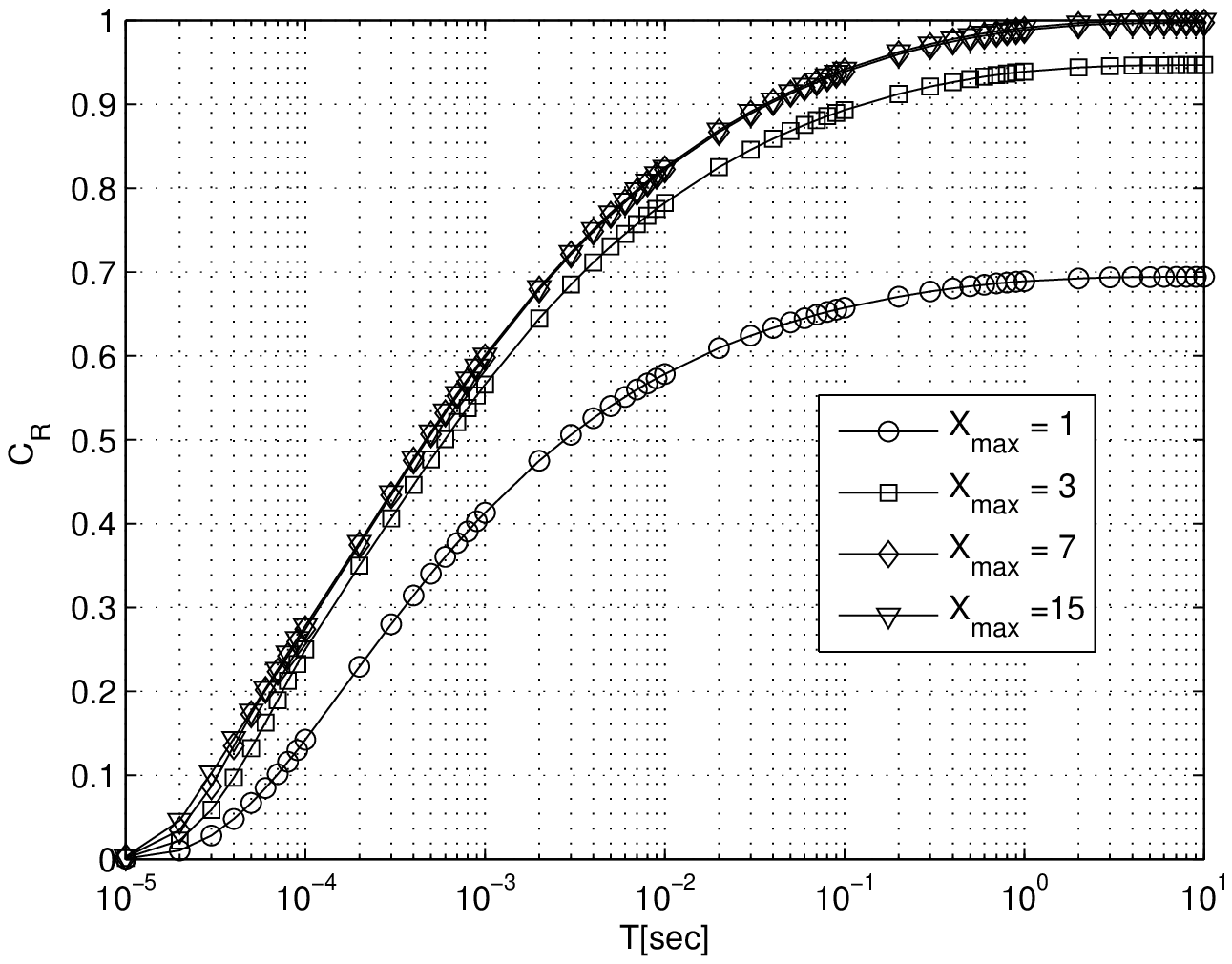}
                \label{fig:C_rel2}
        \end{subfigure}
        \vspace{-1.5\baselineskip}
        \caption{(a) Optimum $a_x$ which maximizes relative capacity for different values of $X_{\max}$ with $T = 10 msec$, $c_0 = 0$. (b) Same as (a) with $c_0 = 1$.(c) $I(X_m;Y_m)$ and (d) $C_R$ in terms of $T$ for different values of $X_{\max}$ with $c_0 = 1$ and all with $\sigma^2 = 1$, $v = 1$, $l = 10^{-2}$.}\label{fig:JKresults}
        \vspace{-1.5\baselineskip}
\end{figure}

Figure~\ref{fig:JKresults} plots the results of the optimization problem in \eqref{eq:SC19}. Figure~\ref{fig:inp_rel_cap} plots the optimum input distribution which maximizes \eqref{eq:SC19A} without shifting the cost function, i.e., letting $c_0 = 0$. As is clear, for all values of $X_{\max}$, the solution is to never transmit any molecules!

The remaining figures in this set use $c_0 = 1$, i.e., $e_x = x + 1$.
Figure~\ref{fig:IN_cos10} plots the optimum input distribution which maximizes \eqref{eq:SC19A} for $T=10$ms. Comparing Figs.~\ref{fig:inp_rel_cap} and~\ref{fig:IN_cos10}, we observe that by changing the cost function to $e(X)= X + 1$, the optimum distribution is radically different, and probability of transmitting non-zero symbols increases (even though $X = 0$ remains the symbol with the highest probability). Figure~\ref{fig:IXY2} plots $I(X_m;Y_m)$ when the input distribution is obtained from \eqref{eq:SC23} for different values of $X_{max}$ with $e(X) = X+1$. Observe that, by increasing $X_{max}$, $I(X_m;Y_m)$ increases. Figure~\ref{fig:C_rel2} plots the relative capacity, $C_R$, for different values of $X_{max}$. We observe that the relative capacity increases as $X_{\max}$ increases, but that quickly saturates to a value of 1 bit/cost.

Table 1 summarizes the effect of parameters in the molecular medium, such as $v$, $l$ and $\sigma^2$, and the transmitter parameters, such as $X_{\max}$ and $E(X)$. Note that the table refers to error rates, an issue considered in Section~\ref{sec:numerical}.
%\begin{table}
% \centering
%             \caption{Summary of results for Capacity Results in DMC.}
%            \includegraphics[width=0.75\textwidth]{Table.eps}
%             \label{fig:Table1}
%             \vspace{-2\baselineskip}
% \end{table}

\begin{table}[t!]
\small
\centering
\caption{Summary of results for Capacity in DMC.}
\label{Table1}
\resizebox{\textwidth}{!}{%
\begin{tabular}{ccccccccc}
\hline
\multicolumn{1}{|c|}{\multirow{2}{*}{\begin{tabular}[c]{@{}c@{}}Objective\\  Function\end{tabular}}}                 & \multicolumn{1}{c|}{\multirow{2}{*}{\begin{tabular}[c]{@{}c@{}}Transmission Cost \\ function\end{tabular}}}                     & \multicolumn{1}{c|}{\multirow{2}{*}{Solution}}                                    & \multicolumn{6}{c|}{Parameters}                                                                                                                                                                                                                                                                                     \\ \cline{4-9}
\multicolumn{1}{|c|}{}                                                                                               & \multicolumn{1}{c|}{}                                                                                                           & \multicolumn{1}{c|}{}                                                             & \multicolumn{1}{c|}{$X_{max}$}                   & \multicolumn{1}{c|}{$\sigma$}                      & \multicolumn{1}{c|}{$v$}                         & \multicolumn{1}{c|}{$l$}                           & \multicolumn{1}{c|}{$T$}                                 & \multicolumn{1}{c|}{$E(X)$}              \\ \hline
\multicolumn{1}{|c|}{\multirow{3}{*}{\begin{tabular}[c]{@{}c@{}} $I\left(X_m;Y_m\right)$\end{tabular}}}                 & \multicolumn{1}{c|}{\multirow{2}{*}{\begin{tabular}[c]{@{}c@{}} $0 \le X_m \le X_{max}$\end{tabular}}} & \multicolumn{1}{c|}{\multirow{2}{*}{\eqref{eq:SC6}}}                              & \multicolumn{1}{c|}{\multirow{3}{*}{$\uparrow$}} & \multicolumn{1}{c|}{\multirow{3}{*}{$ \uparrow $}} & \multicolumn{1}{c|}{\multirow{3}{*}{$\uparrow$}} & \multicolumn{1}{c|}{\multirow{3}{*}{$\downarrow$}} & \multicolumn{1}{c|}{\multirow{3}{*}{$\uparrow$}}         & \multicolumn{1}{c|}{\multirow{2}{*}{NA}} \\
\multicolumn{1}{|c|}{}                                                                                               & \multicolumn{1}{c|}{}                                                                                                           & \multicolumn{1}{c|}{}                                                             & \multicolumn{1}{c|}{}                            & \multicolumn{1}{c|}{}                              & \multicolumn{1}{c|}{}                            & \multicolumn{1}{c|}{}                              & \multicolumn{1}{c|}{}                                    & \multicolumn{1}{c|}{}                    \\ \cline{2-3} \cline{9-9}
\multicolumn{1}{|c|}{}                                                                                               & \multicolumn{1}{c|}{\begin{tabular}[c]{@{}c@{}} $X_m \le X_{max}, E(X_m) \le E$\end{tabular}}      & \multicolumn{1}{c|}{\eqref{eq:SC14}}                                              & \multicolumn{1}{c|}{}                            & \multicolumn{1}{c|}{}                              & \multicolumn{1}{c|}{}                            & \multicolumn{1}{c|}{}                              & \multicolumn{1}{c|}{}                                    & \multicolumn{1}{c|}{$\uparrow$}          \\ \hline
\multicolumn{1}{|c|}{\multirow{2}{*}{\begin{tabular}[c]{@{}c@{}} $I(X_m;Y_m)/T$ \end{tabular}}} & \multicolumn{1}{c|}{\begin{tabular}[c]{@{}c@{}}$0 \le X_m \le X_{max}$\end{tabular}}                  & \multicolumn{1}{c|}{\eqref{eq:SC6}}                                               & \multicolumn{1}{c|}{\multirow{2}{*}{$\uparrow$}} & \multicolumn{1}{c|}{\multirow{2}{*}{$ \uparrow $}} & \multicolumn{1}{c|}{\multirow{2}{*}{$\uparrow$}} & \multicolumn{1}{c|}{\multirow{2}{*}{$\downarrow$}} & \multicolumn{1}{c|}{\multirow{2}{*}{$\nearrow\searrow$}} & \multicolumn{1}{c|}{NA}                  \\ \cline{2-3} \cline{9-9}
\multicolumn{1}{|c|}{}                                                                                               & \multicolumn{1}{c|}{\begin{tabular}[c]{@{}c@{}} $0\le X_m \le X_{max}$,$E(X_m)\le E$ \end{tabular}}        & \multicolumn{1}{c|}{\eqref{eq:SC14}}                                              & \multicolumn{1}{c|}{}                            & \multicolumn{1}{c|}{}                              & \multicolumn{1}{c|}{}                            & \multicolumn{1}{c|}{}                              & \multicolumn{1}{c|}{}                                    & \multicolumn{1}{c|}{$\uparrow$}          \\ \hline
\multicolumn{1}{|c|}{\begin{tabular}[c]{@{}c@{}} ${{I\left( {{X_m};{Y_m}} \right)} \mathord{\left/
 {\vphantom {{I\left( {{X_m};{Y_m}} \right)} {E\left( {{X_m}} \right)}}} \right.
 \kern-\nulldelimiterspace} {E\left( {{X_m}} \right)}}$ \end{tabular}}                  & \multicolumn{1}{c|}{\begin{tabular}[c]{@{}c@{}} \\$0\le X_m\le X_{max}$\end{tabular}}                   & \multicolumn{1}{c|}{\eqref{eq:SC23}}                                              & \multicolumn{1}{c|}{$\downarrow$}                & \multicolumn{1}{c|}{$ \uparrow $}                  & \multicolumn{1}{c|}{$\uparrow$}                  & \multicolumn{1}{c|}{$\downarrow$}                  & \multicolumn{1}{c|}{$\uparrow$}                          & \multicolumn{1}{c|}{NA}                  \\ \hline
\multicolumn{1}{|c|}{\begin{tabular}[c]{@{}c@{}} $P_e$\end{tabular}}                            & \multicolumn{1}{c|}{\begin{tabular}[c]{@{}c@{}} $0\le X_m\le X_{max}$\end{tabular}}                    & \multicolumn{1}{c|}{\eqref{eq:61}}                                                & \multicolumn{1}{c|}{$\uparrow$}                  & \multicolumn{1}{c|}{*}                             & \multicolumn{1}{c|}{$\downarrow$}                & \multicolumn{1}{c|}{$\uparrow$}                    & \multicolumn{1}{c|}{$\downarrow$}                        & \multicolumn{1}{c|}{NA}                  \\ \hline
\multicolumn{1}{|c|}{$T_{opt}$}                                                                                      & \multicolumn{1}{c|}{\begin{tabular}[c]{@{}c@{}}\\$0\le X_m\le X_{max}$\end{tabular}}                  & \multicolumn{1}{c|}{\begin{tabular}[c]{@{}c@{}}Numerical \\ results\end{tabular}} & \multicolumn{1}{c|}{$-$}                  & \multicolumn{1}{c|}{$ \downarrow $}                  & \multicolumn{1}{c|}{$\downarrow$}                & \multicolumn{1}{c|}{$\uparrow$}                    & \multicolumn{1}{c|}{$\uparrow$}                          & \multicolumn{1}{c|}{NA}                  \\ \hline
\multicolumn{9}{l}{\begin{tabular}[c]{@{}l@{}}$ \uparrow $ : Ascending function, $\downarrow$ : Descending function, $ \nearrow\searrow $ : Function have a maximum, \\ * : Depends on $T$, for non large values of $T$ is $\downarrow$, for large values of $T$ is $\uparrow$ \\NA: Not applicable.\end{tabular}}
\end{tabular}
}
\end{table}

\section{Capacity Analysis of Molecular ISI Channel}\label{sec:ISI}

While the previous section analyzed several capacity measures for ASK communications over molecular DMC, in this section, we initiate an analysis in presence of ISI. As stated earlier, we focus on
one time-slot of allowed memory. As with many problems dealing with ISI, we are unable to obtain exact results and resort to bounds. Specifically, we develop two lower bounds and an upper bound for the capacity of the molecular communication system under consideration. We emphasize that the inputs are i.i.d., i.e., the term "capacity" here refers to the simplified case where the inputs within each time-slot are chosen independently and from the same probability distribution. Finally, in this section we develop the maximum a posteriori (MAP) detector (as a possible implementation of a molecular communication receiver) to evaluate the assumption of restricting memory to only one time-slot. We do this by comparing the numerical and simulation results of the performance measure, the probability of error. It is worth noting that one could consider ISI over multiple time-slots using~\eqref{eq:7}. However, this adds exponential complexity to the analysis.

\subsection{Lower bound 1}	
The first lower bound considers the effect of ISI on the mutual information between input and output symbols within the same time-slot. This is a lower bound on the channel capacity because this measure ignores the memory; essentially, we consider a DMC but with an additional source of measurement error due to molecules from the previous time-slot. This lower bound relates to a lower bound of $I_{i.i.d}$ in a discrete-input Gaussian channel with ISI~\cite{10}. Hence we have
\begin{equation}\label{eq:12}
\begin{array}{l}
{I_{L{B_1}}} = I\left( {{X_m};{Y_m}} \right) = H\left( {{Y_m}} \right) - H\left( {\left. {{Y_m}} \right|{X_m}} \right)\\
{\rm{        = }} - \sum\limits_{{y_m} = 0}^{2{X_{\max }}} {p\left( {{y_m}} \right)\log \left( {p\left( {{y_m}} \right)} \right)}  + \sum\limits_{{x_m} = 0}^{{X_{\max }}} {{a_{{x_m}}}} \sum\limits_{{y_m} = 0}^{{x_m} + {X_{\max }}} p \left( {\left. {{y_m}} \right|{x_m}} \right)\log p\left( {\left. {{y_m}} \right|{x_m}} \right)
\end{array}
\end{equation}
where $p\left( {\left. {{y_m}} \right|{x_m}} \right)$ is given by
\begin{equation}\label{eq:13}
p\left( {\left. {{y_m}} \right|{x_m}} \right) = \left\{ {\begin{array}{*{20}{c}}
\begin{array}{l}
{\rm{ }}\sum\limits_{i = 0}^{{x_m}} {\left( {\begin{array}{*{20}{c}}
{{x_m}}\\
i
\end{array}} \right){{\left( {1 - {q_1}} \right)}^{{x_m} - i}}q_1^i\sum\limits_{j = {y_m} - i}^{{X_{\max }}} {\left( {\begin{array}{*{20}{c}}
j\\
{{y_m} - i}
\end{array}} \right){a_j} \times } } {\rm{ }}\\
\hspace*{1in} q_2^{{y_m} - i}{\left( {1 - {q_2}} \right)^{j - \left( {{y_m} - i} \right)}},\begin{array}{*{20}{c}}
{}&{}&{}&{}&{}&{}
\end{array}{\rm{ }}{y_m} \le {x_m} + {X_{\max }}
\end{array}\\
{\begin{array}{*{20}{c}}
0,&{}&{}&{}&{}&{}&{}&{}&{}&{}&{}&{}
\end{array}\hspace*{1.65in} {\rm{}}{y_m} > {x_m} + {X_{\max }}.}
\end{array}} \right.
\end{equation}
By averaging over ${x_m}$ on $p\left( {\left. {{y_m}} \right|{x_m}} \right)$  , $p\left( {{y_m}} \right)$ is given by
\begin{equation}
p\left( {{y_m}} \right) = \sum\limits_{{x_m} = 0}^{{X_{\max }}} {{a_{{x_m}}}\sum\limits_{i = 0}^{{x_m}} {\left( {\begin{array}{*{20}{c}}
{{x_m}}\\
i
\end{array}} \right){{\left( {1 - q} \right)}^{{x_m} - i}}{q^i}} } \sum\limits_{j = y - i}^{{X_{\max }}} {\left( {\begin{array}{*{20}{c}}
j\\
{{y_m} - i}
\end{array}} \right){a_j}} p_2^{{y_m} - i}{\left( {1 - {p_2}} \right)^{j - \left( {{y_m} - i} \right)}}.{\rm{    }}
\end{equation}
Given an input probability vector $\mathbf{a}$, this lower bound can be easily evaluated.
\subsection{Lower Bound 2}
With the one time-slot memory model, the transmitted symbol in the current time-slot only affects the received molecules in the current and the next time-slot. We consider the mutual information between transmitted symbol in current time-slot and received symbols in the current and next time-slot
\begin{eqnarray}\label{eq:14}
{I_{L{B_2}}} & = & I\left( {{X_{m - 1}};{Y_{m - 1}},{Y_m}} \right)
            \mathop  = \limits^{\left( {\rm{a}} \right)} H\left( {{Y_m},{Y_{m - 1}}} \right) - H\left( {\left. {{Y_m},{Y_{m - 1}}} \right|{X_{m - 1}}} \right)\nonumber \\
        & \mathop  = \limits^{\left( {\rm{b}} \right)} & H\left( {{Y_{m - 1}}} \right) + H\left( {\left. {{Y_m}} \right|{Y_{m - 1}}} \right) - H\left( {\left. {{Y_{m - 1}}} \right|{X_{m - 1}}} \right) - H\left( {\left. {{Y_m}} \right|{X_{m - 1}},{Y_{m - 1}}} \right)
\end{eqnarray}
where (a) and (b) are obtained based on the definitions of mutual information and joint entropy, respectively. We consider the channel in the steady state regime, hence $P\left( {\left. {{y_{m - 1}}} \right|{x_{m - 1}}} \right) = p\left( {\left. {{y_m}} \right|{x_m}} \right)$, which is given in \eqref{eq:13}. Also, $p\left( {\left. {{y_m}} \right|{y_{m - 1}},{x_{m - 1}}} \right)$ is given by
\begin{equation}\label{eq:15}
\begin{array}{l}
p\left( {\left. {{y_m}} \right|{y_{m - 1}},{x_{m - 1}}} \right)\mathop  = \limits^{\left( {\rm{a}} \right)} \sum\limits_{{x_m} = 0}^{{X_{\max }}} {p\left( {\left. {{x_m}} \right|{y_{m - 1}},{x_{m - 1}}} \right)p\left( {\left. {{y_m}} \right|{y_{m - 1}},{x_m},{x_{m - 1}}} \right)} \\
\hspace*{0.5in} \begin{array}{*{20}{c}}
{}&{}&{}&{}&{}&{}
\end{array}\mathop  = \limits^{\left( {\rm{b}} \right)} \sum\limits_{{x_m} = 0}^{{X_{\max }}} {p\left( {{x_m}} \right)p\left( {\left. {{y_m}} \right|{y_{m - 1}},{x_m},{x_{m - 1}}} \right)}
\end{array}
\end{equation}
where (a)  is obtained based on the law of total probability, (b) is obtained based on the independence of $x_m$ from $y_{m-1}$ and $x_{m-1}$ which is due to the i.i.d assumption of the input distribution and causality. By averaging over $x_{m-1}$ in \eqref{eq:15}, $p\left( {\left. {{y_m}} \right|{y_{m - 1}}} \right)$ is given by
\begin{equation}\label{eq:16}
\begin{array}{l}
p\left( {\left. {{y_m}} \right|{y_{m - 1}}} \right) = \sum\limits_{{x_{m - 1}} = 0}^{{X_{\max }}} {p\left( {\left. {{x_{m - 1}}} \right|{y_{m - 1}}} \right)p\left( {\left. {{y_m}} \right|{y_{m - 1}},{x_{m - 1}}} \right)} \\
\begin{array}{*{20}{c}}
{}&{}&{}&{}&{}&{}
\end{array}\mathop  = \limits^{\left( {\rm{a}} \right)} \sum\limits_{{x_{m - 1}} = 0}^{{X_{\max }}} {p\left( {\left. {{x_{m - 1}}} \right|{y_{m - 1}}} \right)\sum\limits_{{x_m} = 0}^{{X_{\max }}} {p\left( {{x_m}} \right)} } p\left( {\left. {{y_m}} \right|{y_{m - 1}},{x_m},{x_{m - 1}}} \right)\\
\begin{array}{*{20}{c}}
{}&{}&{}&{}&{}&{}
\end{array}\mathop  = \limits^{\left( {\rm{b}} \right)} \sum\limits_{{x_{m - 1}} = 0}^{{X_{\max }}} {\frac{{p\left( {\left. {{y_{m - 1}}} \right|{x_{m - 1}}} \right)p\left( {{x_{m - 1}}} \right)}}{{p\left( {{y_{m - 1}}} \right)}}} \sum\limits_{{x_m} = 0}^{{X_{\max }}} {p\left( {{x_m}} \right)p\left( {\left. {{y_m}} \right|{y_{m - 1}},{x_m},{x_{m - 1}}} \right)} .
\end{array}
\end{equation}
where (a) and (b) are obtained based on the law of total probability, and Bayes’  rule, respectively. Also, $p\left( {\left. {{y_m}} \right|{y_{m - 1}},{x_m},{x_{m - 1}}} \right)$ is given by
\begin{equation}\label{eq:17}
\begin{array}{l}
p\left( {\left. {{y_m}} \right|{y_{m - 1}},{x_m},{x_{m - 1}}} \right)\mathop  = \limits^{\left( {\rm{a}} \right)} \sum\limits_{{{y'}_{m - 2}} = 0}^{{X_{\max }}} {p\left( {\left. {{{y'}_{m - 2}}} \right|{y_{m - 1}},{x_m},{x_{m - 1}}} \right)} p\left( {\left. {{y_m}} \right|{y_{m - 1}},{x_m},{x_{m - 1}},{{y'}_{m - 2}}} \right)\\
\mathop  = \limits^{\left( {\rm{b}} \right)} \sum\limits_{{{y'}_{m - 2}} = 0}^{{X_{\max }}} {p\left( {\left. {{{y'}_{m - 2}}} \right|{y_{m - 1}},{x_{m - 1}}} \right)} p\left( {\left. {{y_m}} \right|{y_{m - 1}},{x_m},{x_{m - 1}},{{y'}_{m - 2}}} \right)\\
\mathop  = \limits^{\left( {\rm{c}} \right)} \sum\limits_{{{y'}_{m - 2}} = 0}^{{X_{\max }}} {\frac{{p\left( {\left. {{y_{m - 1}},{x_{m - 1}}} \right|{{y'}_{m - 2}}} \right)p\left( {{{y'}_{m - 2}}} \right)}}{{p\left( {{y_{m - 1}},{x_{m - 1}}} \right)}}} p\left( {\left. {{y_m}} \right|{y_{m - 1}},{x_m},{x_{m - 1}},{{y'}_{m - 2}}} \right)\\
\mathop  = \limits^{\left( {\rm{d}} \right)} \sum\limits_{{{y'}_{m - 2}} = 0}^{{X_{\max }}} {\frac{{p\left( {\left. {{y_{m - 1}}} \right|{{y'}_{m - 2}},{x_{m - 1}}} \right)p\left( {\left. {{x_{m - 1}}} \right|{{y'}_{m - 2}}} \right)p\left( {{{y'}_{m - 2}}} \right)}}{{p\left( {\left. {{y_{m - 1}}} \right|{x_{m - 1}}} \right)p\left( {{x_{m - 1}}} \right)}}} p\left( {\left. {{y_m}} \right|{y_{m - 1}},{x_m},{x_{m - 1}},{{y'}_{m - 2}}} \right)\\
\mathop  = \limits^{\left( {\rm{e}} \right)} \sum\limits_{{{y'}_{m - 2}} = 0}^{{X_{\max }}} {\frac{{p\left( {\left. {{y_{m - 1}}} \right|{{y'}_{m - 2}},{x_{m - 1}}} \right)p\left( {{x_{m - 1}}} \right)p\left( {{{y'}_{m - 2}}} \right)}}{{p\left( {\left. {{y_{m - 1}}} \right|{x_{m - 1}}} \right)p\left( {{x_{m - 1}}} \right)}}p\left( {\left. {{y_m}} \right|{y_{m - 1}},{x_m},{x_{m - 1}},{{y'}_{m - 2}}} \right)} \\
 = \sum\limits_{{{y'}_{m - 2}} = 0}^{{X_{\max }}} {\frac{{p\left( {\left. {{y_{m - 1}}} \right|{{y'}_{m - 2}},{x_{m - 1}}} \right)p\left( {{{y'}_{m - 2}}} \right)}}{{p\left( {\left. {{y_{m - 1}}} \right|{x_{m - 1}}} \right)}}p\left( {\left. {{y_m}} \right|{y_{m - 1}},{x_m},{x_{m - 1}},{{y'}_{m - 2}}} \right)} \\
\mathop  = \limits^{\left( {\rm{f}} \right)} \sum\limits_{{{y'}_{m - 2}} = 0}^{{y_{m - 1}}} {\frac{{p\left( {\left. {{{x'}_{m - 1}}} \right|{x_{m - 1}}} \right)p\left( {{{y'}_{m - 2}}} \right)}}{{p\left( {\left. {{y_{m - 1}}} \right|{x_{m - 1}}} \right)}}} p\left( {\left. {{y_m}} \right|{y_{m - 1}},{x_m},{x_{m - 1}},{{y'}_{m - 2}}} \right)
\end{array}
\end{equation}
where (a) is obtained based on the law of total probability and ${y'_{m - 2}}$ is the number of received molecules at the end of time-slot $m-1$ from transmitted molecules in time-slot $m-2$; (b) is obtained due to the independence of ${y'_{m - 2}}$ from $x_m$; (c) is obtained based on Bayes' rule; (d) is obtained based on the joint probability formula; (e) is obtained based on independence of ${y'_{m - 2}}$ and $x_{m-1}$  and the joint probability formula; (f) is obtained due to the fact that ${y_{m - 1}} = {x'_{m - 1}} + {y'_{m - 2}}$, where ${x'_{m - 1}}$ denotes the number of absorbed molecules at time-slot $m-1$ at end of time-slot $m-1$. Based on definition of ${y'_{m - 2}}$ we have
\begin{equation}\label{eq:18}
p\left( {{{y'}_{m - 2}}} \right) = \sum\limits_{{x_{m - 2}} = 0}^{{X_{\max }}} {{a_{{x_{m - 2}}}}\left( {\begin{array}{*{20}{c}}
{{x_{m - 2}}}\\
{{{y'}_{m - 2}}}
\end{array}} \right)} {q_2}^{{{y'}_{m - 2}}}{\left( {1 - {q_2}} \right)^{{x_{m - 2}} - {{y'}_{m - 2}}}}
\end{equation}
\begin{equation}\label{eq:19}
\begin{array}{l}
p\left( {\left. {{y_m}} \right|{y_{m - 1}},{x_m},{x_{m - 1}},{{y'}_{m - 2}}} \right)\mathop  = \limits^{\left( {\rm{a}} \right)} p\left( {\left. {{y_m}} \right|{x_m},{{x''}_{m - 1}} = {x_m} - \left( {{y_{m - 1}} - {{y'}_{m - 2}}} \right)} \right) = \\
\left\{ {\begin{array}{*{20}{c}}
{\sum\limits_{i = {y_m} - {x_{m - 1}}}^{{x_m}} {\left( {\begin{array}{*{20}{c}}
{{x_m}}\\
i
\end{array}} \right){{\left( {1 - {q_1}} \right)}^{{x_m} - i}}{q_1}^i\left( {\begin{array}{*{20}{c}}
{{{x''}_{m - 1}}}\\
{{y_m} - i}
\end{array}} \right)q_2^{{y_m} - i}{{\left( {1 - {q_2}} \right)}^{{{x''}_{m - 1}} - \left( {{y_m} - i} \right)}}{\rm{,}}{y_m} < {x_m} + {{x''}_{m - 1}}} }\\
{{\rm{ }}\begin{array}{*{20}{c}}
0&{}&{}&{}&{}&{}&{}&{}&{}&{}&{}&{}&{}&{}&{}&{}
\end{array}{\rm{,}}{y_m} > {x_m} + {{x''}_{m - 1}}}
\end{array}} \right.
\end{array}
\end{equation}
(a) is obtained because ${x_{m - 1}} = {x'_{m - 1}} + {x''_{m - 1}}$, where ${x''_{m - 1}}$ is the number of remaining molecules at end of time-slot $m-1$   from transmitted molecules in the same time-slot. Moreover, based on definition of ${x'_{m - 1}}$ , $P\left( {\left. {{{x'}_{m - 1}}} \right|{x_{m - 1}}} \right)$ is given by
\begin{equation}\label{eq:20}
P\left( {\left. {{{x'}_{m - 1}}} \right|{x_{m - 1}}} \right) = \left( {\begin{array}{*{20}{c}}
{{x_{m - 1}}}\\
{{{x'}_{m - 1}}}
\end{array}} \right)q_1^{{{x'}_{m - 1}}}{\left( {1 - {q_1}} \right)^{{x_{m - 1}} - {{x'}_{m - 1}}}}.
\end{equation}
While far more involved, as with the first lower bound in~\eqref{eq:12}, given the input probability distribution, obtaining this bound is a simple matter.

\subsection{Upper Bound}
Having developed two lower bounds on capacity, we now develop an upper bound motivated by the matched filter upper bound in Gaussian ISI channels~\cite{10}. With one time-slot memory for the channel, if we transmit symbols and wait two time-slots before transmitting the next symbol and remove any subsequently arriving molecules, we have an interference-free channel and arrive at an upper bound to the capacity per channel use. This is equivalent to the DMC case with the binomial transition probabilities of~\eqref{eq:2} where $q_1$ is replaced by $q_U = F_W(2T)$. The mutual information of this channel is concave which can be maximized using the Blahut-Arimoto algorithm~\cite{17}. We therefore have the upper bound
\begin{equation}\label{eq:21}
{I_{UB}} = I\left( {{X_m};{Y_m}} \right) = H\left( {{Y_m}} \right) - H\left( {\left. {{Y_m}} \right|{X_m}} \right)
\end{equation}
where $P\left( {\left. {{y_m}} \right|{x_m}} \right)$ is given by
\begin{equation}\label{eq:22}
P\left( {\left. {{y_m}} \right|{x_m}} \right) = \left\{ {\begin{array}{*{20}{c}}
{\left( {\begin{array}{*{20}{c}}
{{x_m}}\\
{{y_m}}
\end{array}} \right)q_U^{^{{y_m}}}{{\left( {1 - {q_U}} \right)}^{{x_m} - {y_m}}}{\rm{,}}{y_m} \le {x_m}}\\
{\begin{array}{*{20}{c}}
0&{}&{}&{}&{}&{}&{}
\end{array}{\rm{ ,}}{y_m} > {x_m}.}
\end{array}} \right.
\end{equation}
and ${q_U} = {F_W}\left( {2T} \right)$.
Clearly,
\begin{equation}\label{eq:27}
\begin{array}{l}
{I_{L{B_2}}} = I\left( {{X_{m - 1}};{Y_{m - 1}},{Y_m}} \right)\mathop  = \limits^{\left( {\rm{a}} \right)} I\left( {{X_{m - 1}};{Y_{m - 1}}} \right) + I\left( {{X_{m - 1}};{Y_m}\left| {{Y_{m - 1}}} \right.} \right)\\
\begin{array}{*{20}{c}}
{}&{}&{}&{}&{}&{}&{}&{}
\end{array}\mathop  \ge \limits^{\left( b \right)} I\left( {{X_{m - 1}};{Y_{m - 1}}} \right) = {I_{L{B_1}}}
\end{array}
\end{equation}
where (a) is obtained from chain rule in mutual information and (b) is obtained based on the non-negativity of mutual information. Hence, we have the following result
\[{I_{L{B_1}}} \le {I_{L{B_2}}} \le {I_{i.i.d}} \le {I_{UB}}, \]
as long as the input distribution for the upper bound is optimum. We discuss optimizing the bounds next.
\subsection{Optimizing Lower and Upper Bounds}
We are unable to show the concavity of the lower bounds ${I_{L{B_1}}}$  and ${I_{L{B_2}}}$  with respect to the input distribution vector $\mathbf{a}$; however, we can modify the Blahut-Arimoto algorithm~\cite{17} to find a local maximum, and as a result, we can optimize, to within a local maximum, these two lower bounds. The channel transition matrices ${{\bf{P}}^{\left( {L{B_h}} \right)}}$, $h \in \left\{ {1,2} \right\}$ for lower bounds 1 and 2 are
\begin{equation}\label{eq:29}
{{\bf{P}}_{x_{m},y_{m}}^{\left( {L{B_1}} \right)}} = \left[ {p\left( {\left. {{Y_m} = {y_m}} \right|{X_{m}} = {x_m}} \right)} \right]
\end{equation}
\begin{equation}\label{eq:30}
{{\bf{P}}_{x_{m-1},y_{m-1}y_m}^{\left( {L{B_2}} \right)}} = \left[ {p\left( {\left. {{Y_m} = {y_m},{Y_{m - 1}} = {y_{m - 1}}} \right|{X_{m - 1}} = {x_{m - 1}}} \right)} \right]
\end{equation}
where, importantly, the size of ${{\bf{P}}^{\left( {{LB}_1} \right)}}$  and ${{\bf{P}}^{\left( {{LB}_2} \right)}}$  are  $\left( {{X_{\max }} + 1} \right) \times \left( {2{X_{\max }} + 1} \right)$ and  $\left( {{X_{\max }} + 1} \right) \times {\left( {2{X_{\max }} + 1} \right)^2 }$ respectively. Also, ${y_{m,}}{y_{m - 1}} \in \left[ {0,...,2{X_{\max }}} \right]$   and ${x_m} \in \left[ {0,...,{X_{\max }}} \right]$ . For matrix ${\bf{Q}}$ with size of ${{\bf{P}}^{\left( {{LB}_h} \right)}}$, $h \in \left\{ {1,2} \right\}$ , let
\begin{equation}\label{eq:31}
J\left( {{\bf{a}},{{\bf{P}}^{\left({LB}_h\right)}},{\bf{Q}}} \right) = \sum\nolimits_j {\sum\nolimits_i {{a_j}{\bf{P}}_{j,i}^{\left({LB}_h\right)}\log \frac{{{{\bf{Q}}_{i,j}}}}{{{a_j}}}} } .
\end{equation}
Then the following is true.
 \begin{enumerate}
   \item $I_{{LB}_{h}} = \mathop {\max }\limits_{\bf{a}} \mathop {\max }\limits_{\bf{Q}} J\left( {{\bf{a}},{\bf{P}}_{}^{\left({LB}_h\right)},{\bf{Q}}} \right).$
   \item For fixed ${\bf{a}}$, $J\left( {{\bf{a}},{\bf{P}}^{\left({LB}_h\right)},{\bf{Q}}} \right)$ is locally maximized by
   \begin{equation}\label{eq:32}
   {{\bf{Q}}_{k,j}} = \frac{{{a_j}{\bf{P}}_{j,k}^{L{B^{\left( h \right)}}}}}{{\sum\nolimits_j {{a_j}{\bf{P}}_{j,k}^{L{B^{\left( h \right)}}}} }}.
   \end{equation}
   \item For fixed ${\bf{Q}}$, $J\left( {{\bf{a}},{\bf{P}}_{}^{\left({LB}_h\right)},{\bf{Q}}} \right)$  is locally maximized by
   \begin{equation}\label{eq:33}
      {a_j} = \frac{{\exp \left( {\sum\nolimits_i {{\bf{P}}_{j,i}^{\left({LB}_{h}\right)}\log {{\bf{Q}}_{i,j}}} } \right)}}{{\sum\nolimits_j {\exp \left( {\sum\nolimits_i {{\bf{P}}_{j,i}^{\left({LB}_h\right)}\log {{\bf{Q}}_{i,j}}} } \right)} }}
   \end{equation}
 \end{enumerate}
The algorithm iterates between $a_j$ derived in \eqref{eq:33} and the transition probability matrices ${{\bf{P}}^{\left( {L{B_h}} \right)}}$ in \eqref{eq:29} or \eqref{eq:30}. This procedure is repeated until the convergence of the probabilities ${a_j}$.

In contrast to the lower bounds, ${I_{UB}}$ is a concave function in terms of ${\textbf{a}}$. Hence, using the standard Blahut-Arimoto algorithm~\cite{17} a unique distribution globally maximizing the upper bound can be obtained (note that this is critical for ${I_{UB}}$ to be a valid upper bound).

\subsection{MAP and ML Detectors}

In this subsection, we derive the detection performance of a molecular ASK receiver. Since the true channel capacity is unknown, we are unable to directly evaluate the quality of the assumption of one time-slot memory in molecular ISI channel on the capacity analysis. As a proxy, we examine the effect of this assumption on the detection performance using simulations and analysis. In general, one may resort to a sequence detector over an ISI channel for improved performance~\cite{AkanRec, NoelRec, ShahRec}, however, in molecular communications the implementation complexity for nano-machines is likely to be prohibitive. Hence, here we consider a MAP symbol-by-symbol detector at the receiver to estimate the number of molecules transmitted. Hence, here we consider a MAP symbol-by-symbol detector at the receiver to estimate the number of molecules transmitted. The decision rule is given by
\begin{equation}\label{eq:59}
\begin{array}{l}
{{\hat x}} = \arg \mathop {\max }\limits_x {p_{\left. {{Y_k}} \right|{X_k}}}\left( {\left. x \right|y} \right)=  \arg \mathop {\max }\limits_x {p_{\left. {{Y_k}} \right|{X_k}}}\left( {\left. y \right|x} \right)p\left( x \right)
\end{array}
\end{equation}
By replacing \eqref{eq:6} in \eqref{eq:59}, we have
\begin{equation}\label{eq:60}
\hat x = \arg \mathop {\max }\limits_x \sum\nolimits_{i = 0}^{y - x} {{a_x}\left( {\begin{array}{*{20}{c}}
x\\
i
\end{array}} \right)q_1^i{{\left( {1 - {q_1}} \right)}^{x - i}}{P_k}\left( {y - i} \right)}
\end{equation}
If $x$ molecules are transmitted, the decision region for detecting $x$ at the receiver is denoted by $\left\{ {{Y_x}} \right\}$. Hence, probability of error is denoted by
\begin{equation}\label{eq:61}
{P_e} = \sum\limits_{x = 0}^{{X_{\max }}} {P\left( {\left. {y \notin \left\{ {{Y_x}} \right\}} \right|x} \right)a_x}
\end{equation}
As always, using a uniform input distribution in $a_x$, makes the MAP detector a ML detector.

\subsection{Numerical Results} \label{sec:numerical}
\begin{figure}
    \centering
        \vspace{-1.5\baselineskip}
        \begin{subfigure}[b]{0.49\textwidth}
                \caption{}
                \includegraphics[width=\textwidth]{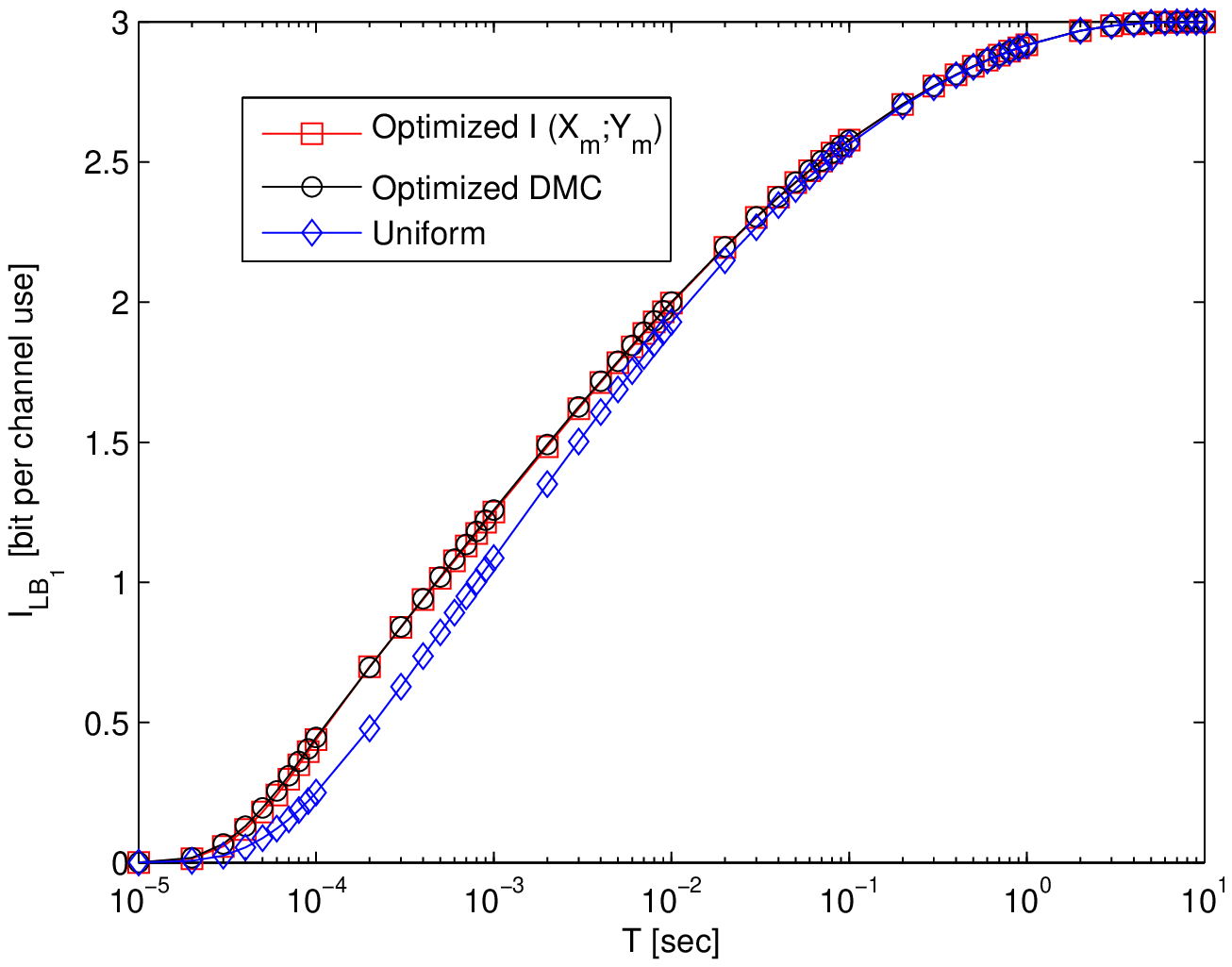}
                \label{fig:I_LB1}
        \end{subfigure}%
        \begin{subfigure}[b]{0.49\textwidth}
                \caption{}
                \includegraphics[width=\textwidth]{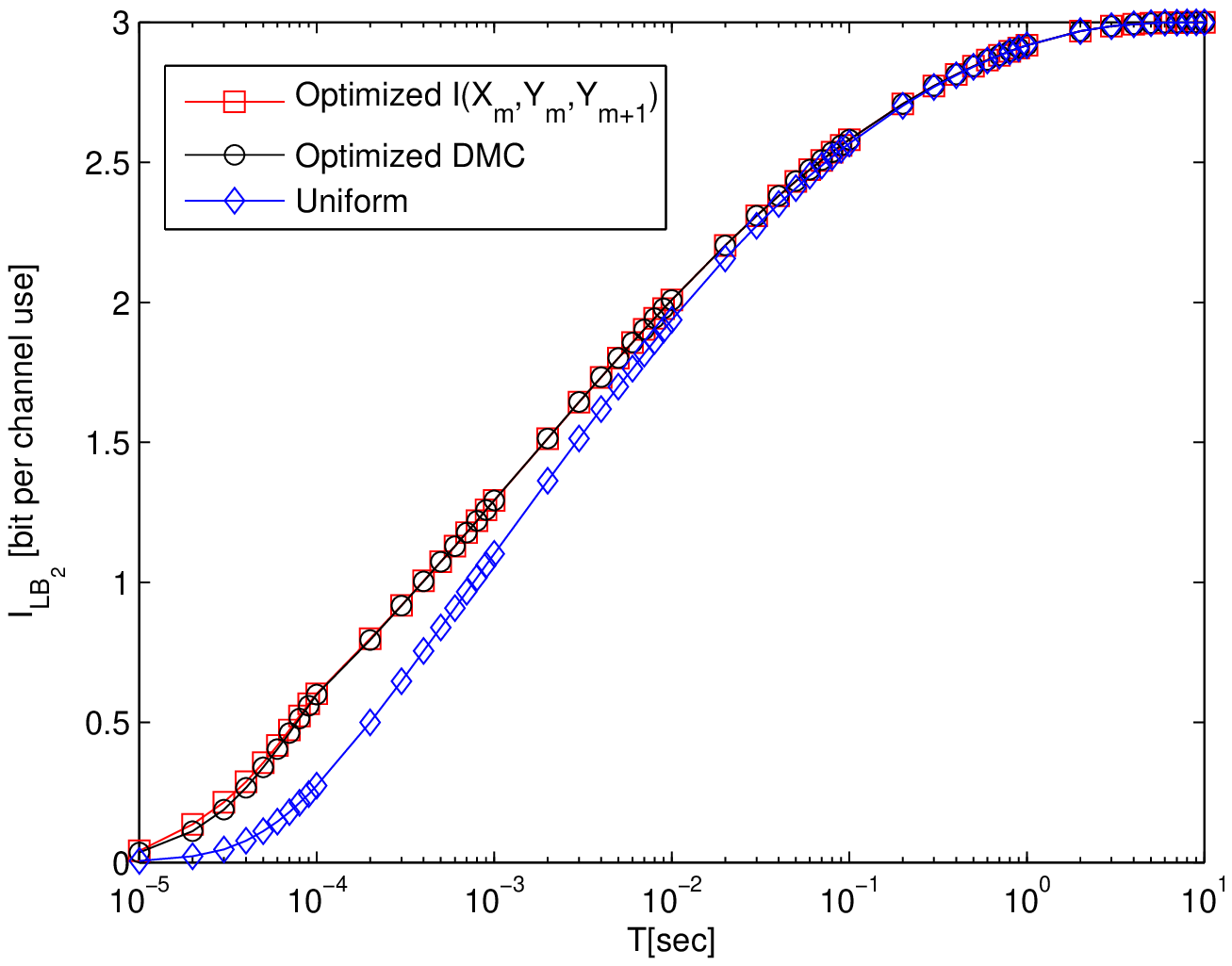}
                \label{fig:I_LB2}
        \end{subfigure}\\%
        \vspace{-1.5\baselineskip}
        \begin{subfigure}[b]{0.49\textwidth}
                \caption{}
                \includegraphics[width=\textwidth]{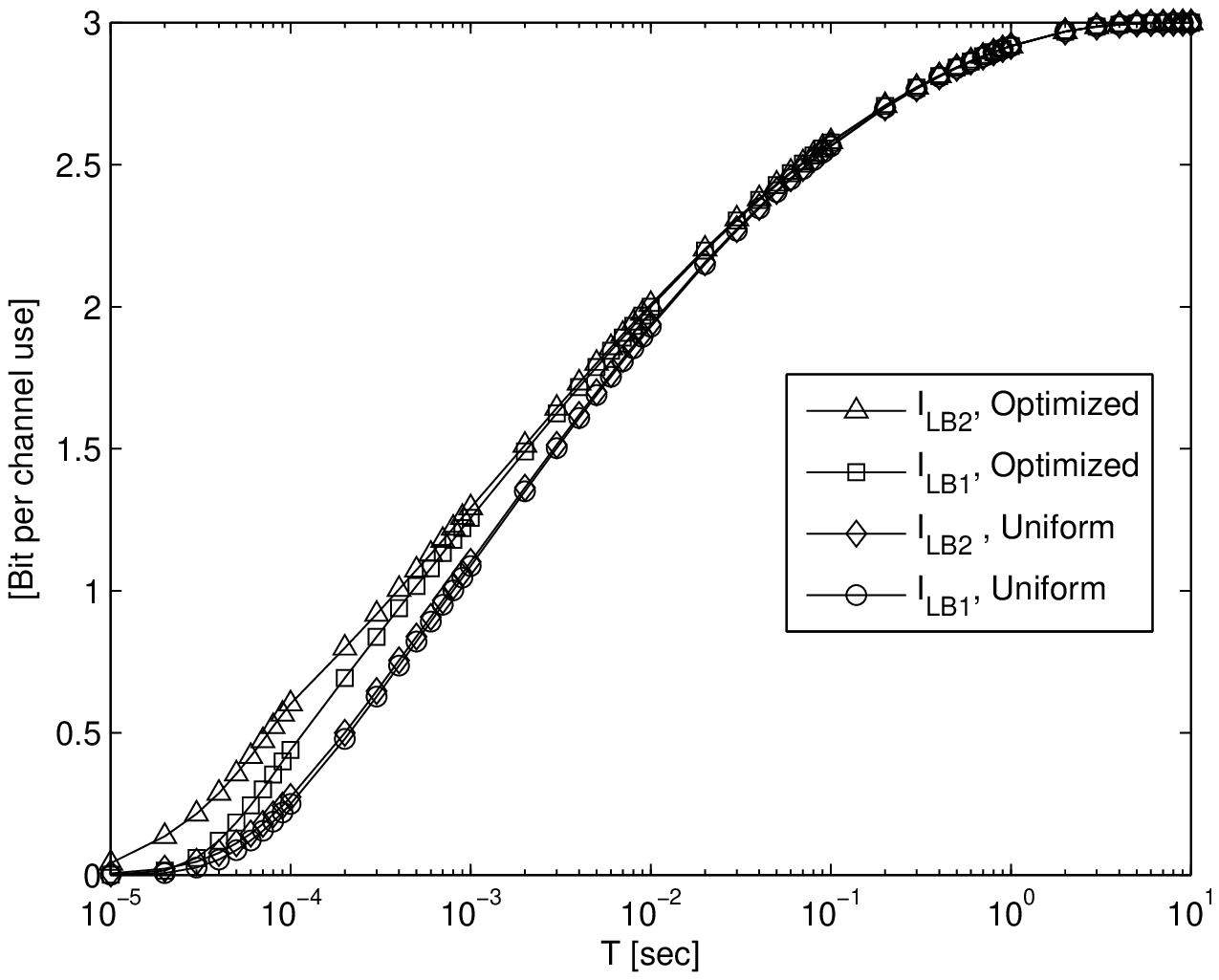}
                \label{fig:I_LB}
        \end{subfigure}
        \vspace{-1.5\baselineskip}
        \caption{(a-b) Two lower bounds versus $T$ (c) Comparing the lower bounds for uniform and optimized input distributions. $v = 1$, $l = 10^{-2}$, $\sigma^2 = 1$, $X_{max} = 7$.} \label{fig:LB1LB2}
\end{figure}

We now evaluate and compare the derived bounds for various medium parameters such as $l$, $v$ and $\sigma$. Figures~\ref{fig:I_LB1} and~\ref{fig:I_LB2} plot the two lower bounds. We quantify the lower bounds for three input distributions: (i) optimized input distribution from maximizing $I(X_m;Y_m)$ ($I(X_m;Y_m,Y_{m+1})$), denoted by "Optimized $I(X_m,Y_m)$" ("Optimized $I(X_m;Y_m,Y_{m+1})$ ") in the legend for $I_{LB_1}$ ($I_{LB_2}$), (ii) optimized input distribution from maximizing the mutual information of DMC in~\eqref{eq:SC1}, denoted by "Optimized DMC" in the legend, and (iii) uniform input distribution. This study shows the impact of selecting input distribution on the numerical value of capacity bounds.
One sees that using the optimized input distributions, the bounds ${I_{L{B_1}}}$ and ${I_{L{B_2}}}$ improve noticeably in comparison to the cases with uniform input distribution. This is while a uniform distribution maximizes the source entropy and the capacity of a molecular channel without ISI (here an error free channel). Another important observation here is that each of the two proposed bounds for the molecular ISI channel amounts to almost an equal rate with either its corresponding optimized distribution or that from the molecular DMC case in Section~\ref{sec:DMC}. To directly compare the two bounds, Figure~\ref{fig:I_LB} compares ${I_{{LB}_1}}$ and ${I_{{LB}_2}}$  as a function of $T$ for the optimized input distribution from~\eqref{eq:33} (labeled by "Optimized"), and a uniform input distribution.

%%Figure 6
%\begin{figure}
%        \centering
%        \includegraphics[width=0.44\textwidth]{fig_4.eps}
%        \caption{Comparing the lower bounds for uniform and optimized input distributions. $v = 1$, $l = 10^{-2}$, $\sigma^2 = 1$, $X_{max} = 7$.}
%        \label{fig:I_LB}
%\end{figure}

\begin{figure}
        \centering
        \vspace{-1.5\baselineskip}
        \begin{subfigure}[h!]{0.49\textwidth}
                \caption{}
                \includegraphics[width=\textwidth]{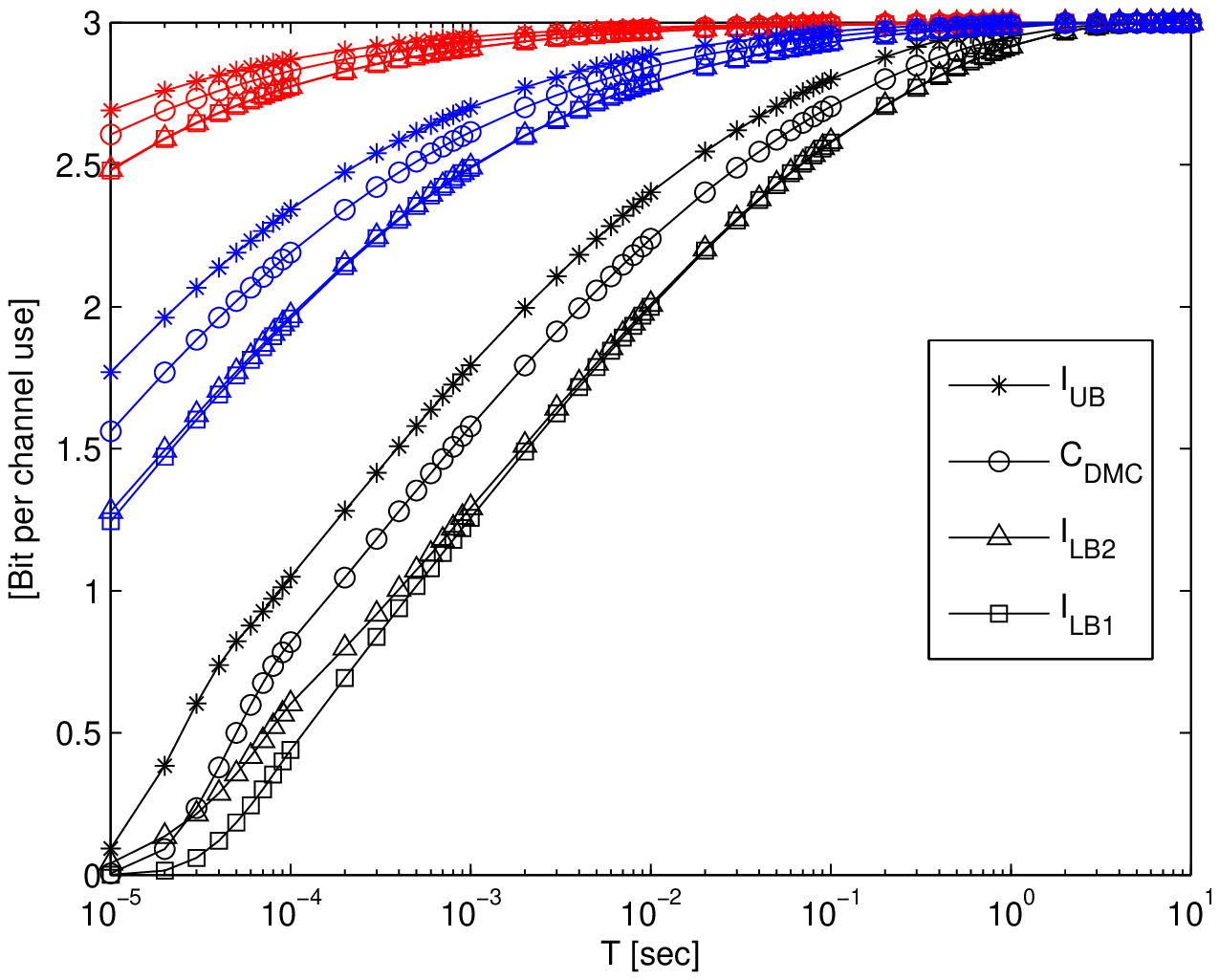}
                \label{fig:I_l}
        \end{subfigure}%
         \begin{subfigure}[h!]{0.49\textwidth}
                \caption{}
                \includegraphics[width=\textwidth]{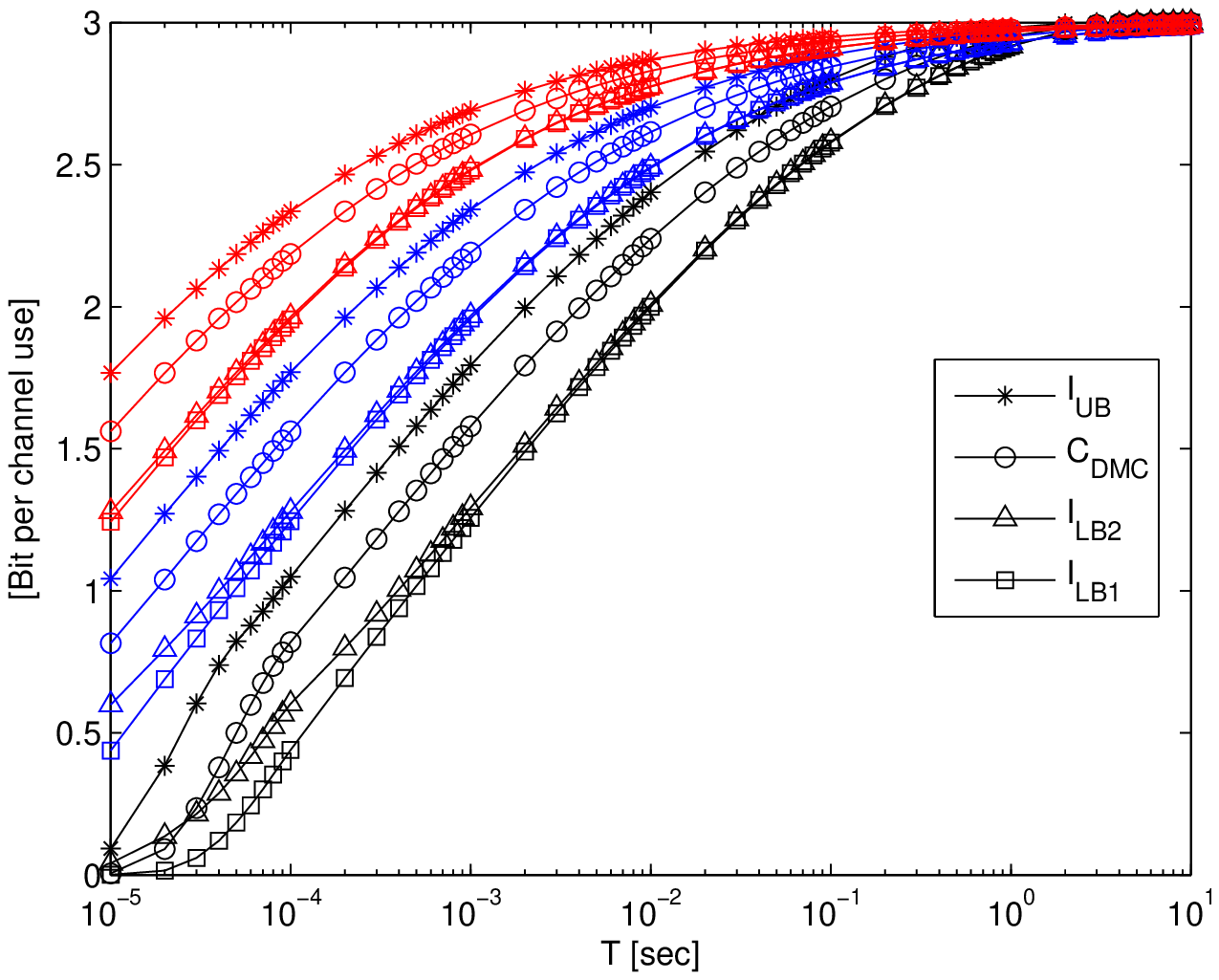}
                \label{fig:I_sig}
        \end{subfigure}
        \\
        \vspace{-1.5\baselineskip}
        \begin{subfigure}[h!]{0.49\textwidth}
                \caption{}
                \includegraphics[width=\textwidth]{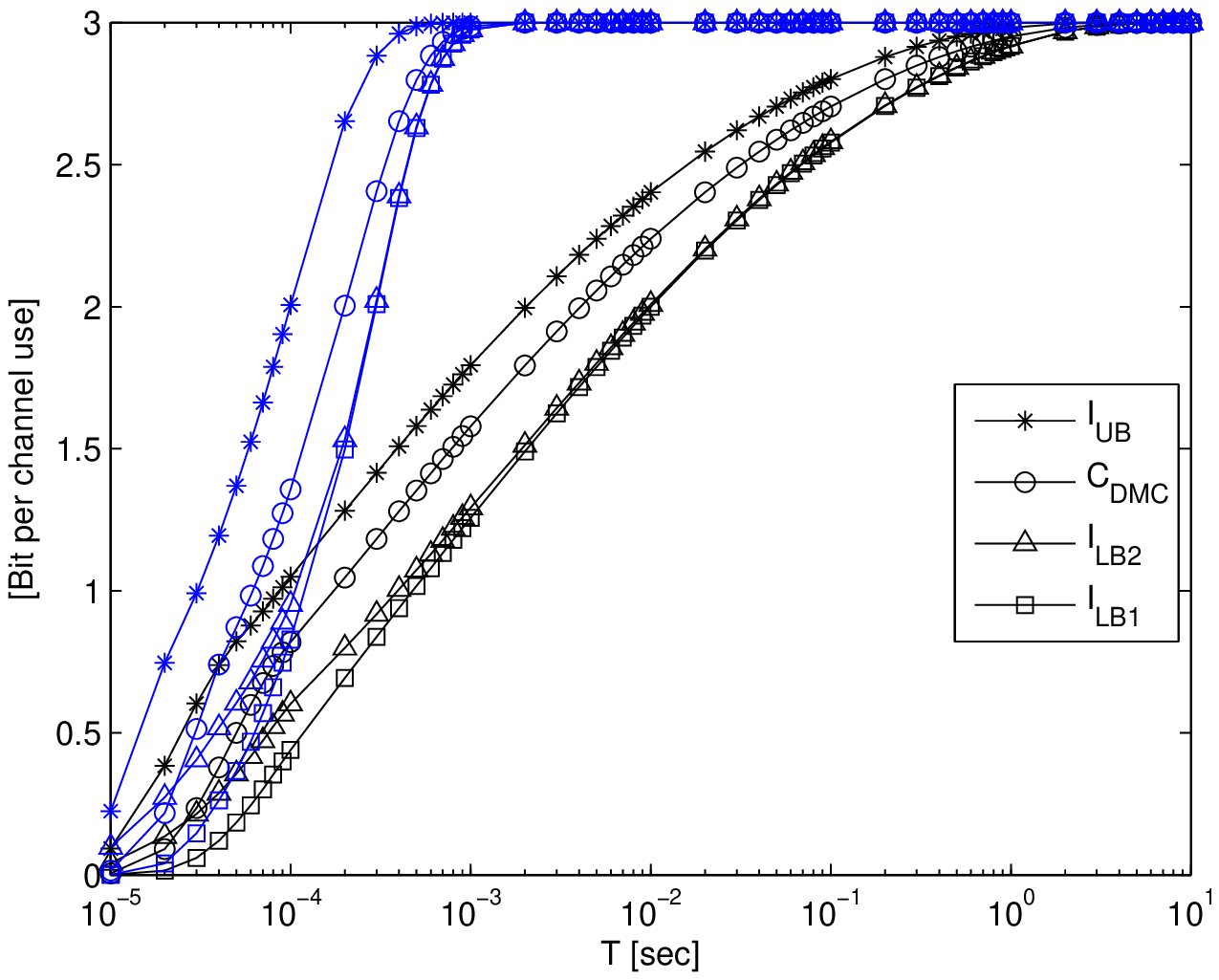}
                \label{fig:I_v}
        \end{subfigure}%
        \vspace{-1.5\baselineskip}
        \caption { ${I_{L{B_2}}}$, ${I_{L{B_1}}}$, ${C_{DMC}}$ and ${I_{UB}}$ in terms of $T$ for different values of (a) $l$ (Black curves, $l = 10^{-2}$, blue curves, $l = 10^{-3}$, red curves, $l = 10^{-4}$) with $v = 1$, $\sigma^2 = 1$, (b) $\sigma^2$ (Black curves, $\sigma^2 = 1$, blue curves, $\sigma^2 = 10$, red curves, $\sigma^2 = 100$) with $v = 1$, $\l = 10^{-2}$, (c) $v$ (Black curves, $v = 1$, blue curves, $v = 100$) with $\l = 10^{-2}$, $\sigma^2 = 1$ all with $X_{max} = 7$.} \label{planBexhas}
\end{figure}

Figures~\ref{fig:I_l}-\ref{fig:I_v} plot ${I_{{LB}_2}}$, ${I_{{LB}_1}}$, ${C_{DMC}}$ and ${I_{UB}}$ versus $T$ for different values of the transmitter-receiver distance, $l$, the diffusion constant $\sigma$ and drift velocity $v$, respectively. As evident, by decreasing $l$ all bounds increase and converge to ${\log _2}\left( {{X_{\max }} + 1} \right) = 3$, which is the entropy of the source. Similarly, all capacity bounds are increasing functions of $\sigma$. As also observed in~\cite{2} (for the case when information is encoded in the time of release) increasing drift velocity increases mutual information due to reduced ISI. Comparing the results of the three recent sub-figures, we note that the bounds are most sensitive to the transmitter-receiver distance $l$ and drift velocity $v$ while not being as sensitive to the diffusion constant $\sigma$. Crucially, over wide ranges of $T$, the upper and lower bounds are close. Also, due to the reduced ISI, reducing $l$ reduces the gap between all the derived bounds.

The capacity of an ASK-based molecular communication channel is in general unknown. To test the validity of the one time-slot memory assumption, we compare the probability of error of the ML detector in three cases: no ISI (the DMC of Section~\ref{sec:DMC}), including ISI up to one time-slot (the case analyzed in Section~\ref{sec:ISI}) and simulation results that track the arrivals of all molecules. We reason that, if our assumption is invalid, there should be a noticeable difference between the error rates for the three cases. In all cases, we use a uniform input distribution (making the MAP and ML detectors equivalent).

\begin{figure}
        \centering
        \vspace{-1\baselineskip}
        \begin{subfigure}[h!]{0.75\textwidth}
         \vspace{-1.5\baselineskip}
                \caption{}
                \includegraphics[width=\textwidth]{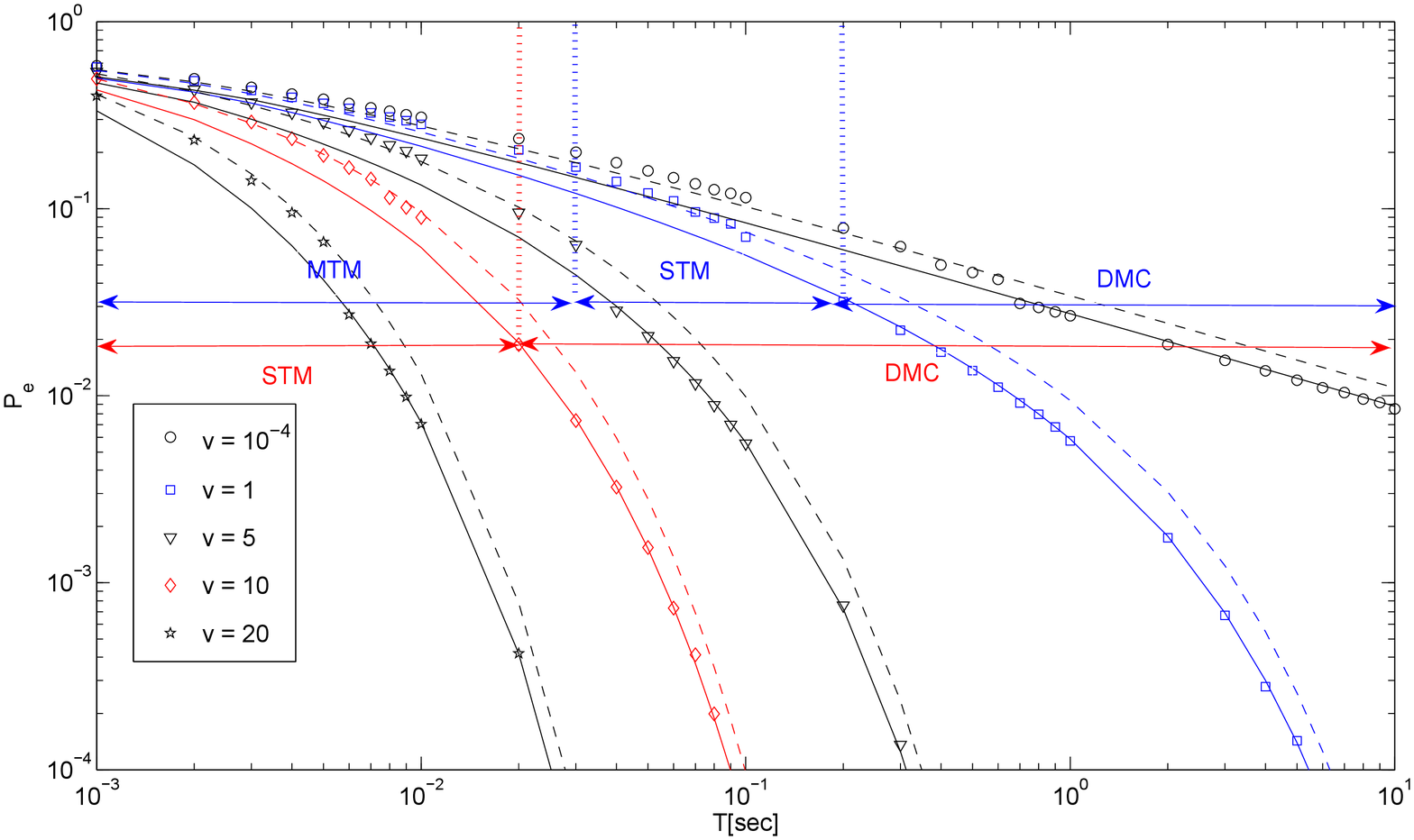}
                \label{fig:P_e1}
        \end{subfigure}\\%
 %       \vspace{-1.5\baselineskip}
        \begin{subfigure}[b]{0.75\textwidth}
        \vspace{-1\baselineskip}
                \caption{}
                \includegraphics[width=\textwidth]{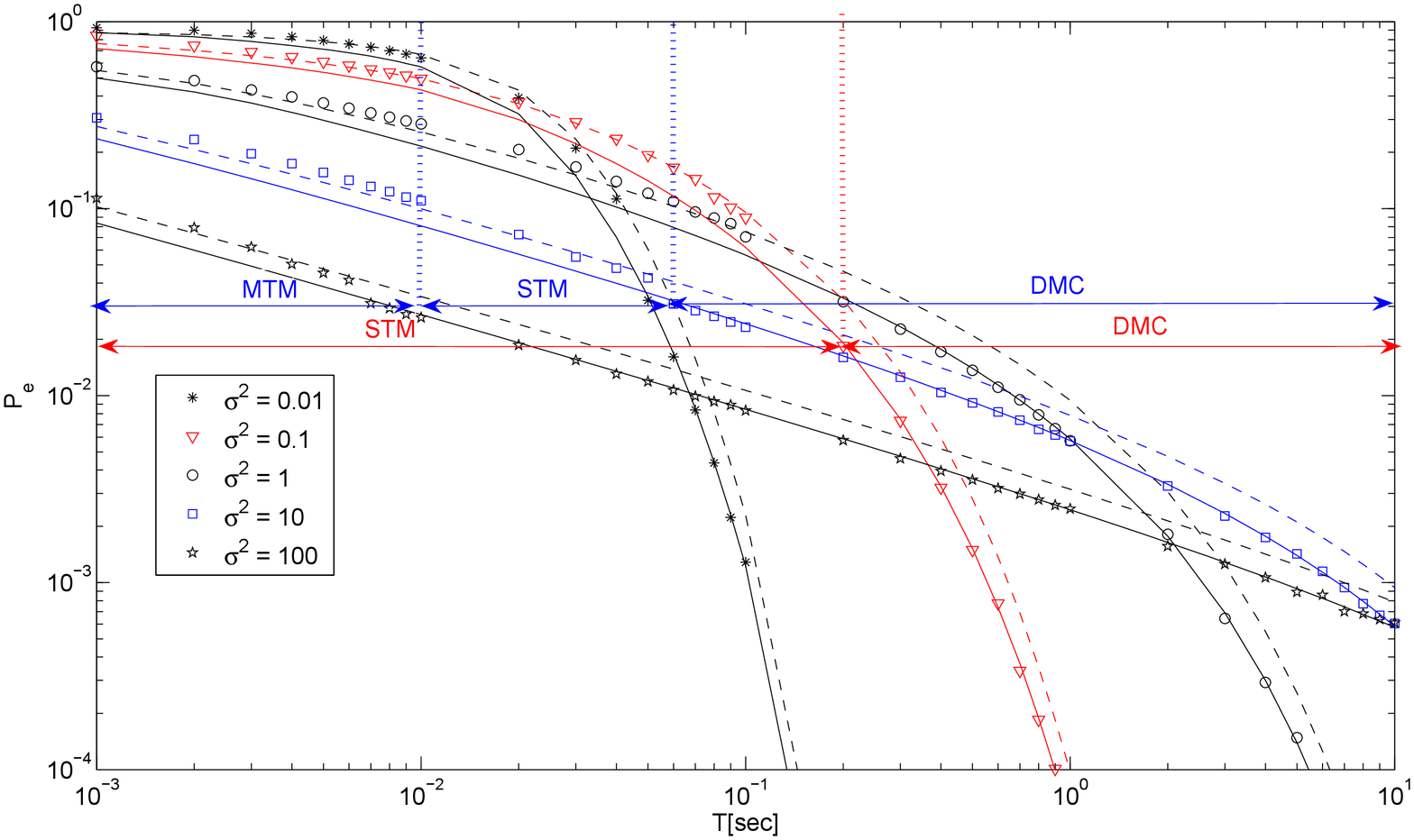}
                \label{fig:P_e2}
        \end{subfigure}%
        \vspace{-1.5\baselineskip}
%        \begin{subfigure}[b]{0.49\textwidth}
%                \caption{}
%                \includegraphics[width=\textwidth]{fig_9.eps}
%                \label{fig:P_e2}
%        \end{subfigure}%
%        \vspace{-1.5\baselineskip}
%        \begin{subfigure}[b]{0.49\textwidth}
%                \caption{}
%                \includegraphics[width=\textwidth]{fig_10.eps}
%                \label{fig:P_e3}
%        \end{subfigure}%

        \caption { $P_e$ in terms of $T$ using ML detector for different values of (a) $v$  with $l = 10^{-2}$, $\sigma^2 = 1$ (b) $\sigma^2$  with $l = 10^{-2}$, $v = 1$, all with $X_{\max} = 7$. (Solid lines: numerical results of calculated $P_e$ over DMC, Dashed lines: numerical results of calculated $P_e$ with single time-slot memory (STM), markers: Simulation results (multiple time-slot memory (MTM)). % (d) $P_e$  of Viterbi detector and $P_e$  of \eqref{eq:60} in terms of $T$ for MAP detector and different values of $X_{max}$ with $l = 10^-2$, $v=1$, $\sigma^2 = 1$.
        )} \label{fig:P_e_sim}
\end{figure}

Figure~\ref{fig:P_e_sim} plots the probability of error, $P_e$, of the ML detector versus $T$. In this figure, (\emph{i}) the solid curves denote the error probability for the molecular DMC case, (\emph{ii}) the dashed curves denote the error assuming single time-slot memory (STM) ISI model, and (\emph{iii}) the markers denote simulation results based on tracking all molecules (labeled as the multiple time-slot ISI model (MTM)). By comparing whether the markers match the dashed or solid curves in these figures, one can examine the range of parameters in which each of the DMC, STM or MTM models is valid. In Fig.~\ref{fig:P_e1}, $P_e$ is depicted versus $T$ for different values of $v$ with $l = 10^{-2}$ and $\sigma^2 = 1$. One sees that with $v = 10$, for $T > 2 \times 10 ^{-2}$, the numerical results of DMC match the simulation results and for $3\times10^{-3}<T< 2 \times 10^{-2}$, numerical results of STM match the simulation results; hence for $v = 10$, the DMC and STM models are valid in the said ranges, respectively. Also, Figure \ref{fig:P_e2} shows $P_e$ in terms of $T$ for different values of $\sigma^2$ with $v=1$ and $l=10^{-2}$. We observe that for small to medium values of  $T$, increasing $\sigma^2$ reduces $P_e$, while for large values of $T$, increasing $\sigma^2$, increases $P_e$ instead. This is consistent with the behavior of $q_1$ as a function of $\sigma^2$. For small and medium values of $T$, in which ISI is important, $q_1$ is an increasing function in terms of $\sigma^2$, i.e., the increased variance in position increases the probability of the molecules arriving within two time-slots after release. Moreover, one sees in this figure that for $\sigma^2 = 0.1$, the DMC and STM models are valid for $T > 3 \times 10^{-1}$ and $10^{-3}< T <3 \times 10^{-1}$, respectively.

It is evident in Figure~\ref{fig:P_e_sim}, that the numerical results of the DMC case (the solid curves) match well to the simulation results for higher values of $T$, while those of the one time-slot memory model (the dashed curves) match the simulation results for lower values of $T$. The results show that accounting for one time-slot of memory is adequate for design and capacity analysis purposes in an interesting range of molecular channel parameters and error probabilities. Specifically, the corresponding results are very close to the more accurate case of memory spread over multiple time-slots. For more accurate analysis in these cases, one may opt to extend the presented analysis to the case where molecules arriving with more than one time-slot delay are also taken into account in modeling ISI. Naturally, this is only achieved at the cost of substantial (exponentially growing) computational complexity in the analysis.

For small values of $T$ the duration of ISI is more than one time-slot, and the probability of error performance of the ML detector deteriorates. This is naturally assuming that the delayed molecules do not expire over time. As stated, one may resort to more advanced and complex sequence detectors over such harsh ISI channels to increase the rate and yet maintain a desired error performance. The analysis and design of such schemes may be investigated in future studies.

It is noteworthy that depending on the characteristics of the fluid media, the quantity of molecules and their life expectancy, the slot duration should be chosen such that a statistical majority of molecules emitted in a slot arrive within a time-slot. Of course, the presented model, detectors and analyses will serve most effectively in such a practical setting.

%figure 8

\clearpage
\section{Conclusions}
In this paper, we analyze the capacity of a molecular communications channel when information is encoded in the number of transmitted molecules (molecular ASK). The molecules propagation is governed by Brownian motion and the probability of arrival within a specific time-slot is derived using the additive inverse Gaussian model for the transmission time. We analyzed the capacity in both DMC and ISI cases.

For the DMC, the optimized input distribution was derived to achieve the capacity per channel use, per unit time and per unit cost. Our results show that for small (large) values of transmission time-slots, the distribution achieving the capacity per channel use converges to bipolar (uniform) distribution. Also, imposing a limitation on average number of transmitted molecules per channel use reduces the capacity per channel use and makes symbols with lower transmission cost more probable. As a result in this case, the optimum input distribution, maximizing the capacity per channel use, deviates from a uniform distribution even for large values of time-slot duration.
By considering the capacity per unit time as the objective function, the optimum time-slot duration, which maximizes the capacity with peak and/or average constraints on the number of transmitted molecules per unit time, was obtained numerically. We studied the dependence of the optimum time-slot duration with parameters of the molecular medium. This paper also offered the first study on the capacity per unit cost in molecular communications, which is a capacity measure conscious of its relative efficiency with respect to the molecular injection rate. For the case where molecular alphabet only consists of symbols with non-zero cost, a non-zero capacity per unit cost is feasible and we obtained the corresponding optimized input distribution.

The second half of this paper analyzes the case with ISI; specifically, with ISI restricted to one time-slot. Two lower bounds and an upper bound for ASK-based molecular communication with ISI were derived. Our results quantified how the lower bounds improve when the corresponding input symbol distributions are optimized. Importantly, our results showed that over a wide range of parameter values the gap between the lower and upper bounds are small and so can provide a good measure of capacity. The results also showed that the optimum distribution obtained assuming a DMC provides close-to-optimal results for the ISI case as well.

To test the validity of the one time-slot memory model for the molecular ISI channel, we compared the performance of the ML detector in this case, with that assuming a DMC model and the results from the simulations. As our results show, the one time-slot memory model is valid over an interesting range of molecular channel parameters. One may extend the presented analysis to the case with multiple time-slot memory models at the cost of increased computational complexity. Also, the design and analysis of efficient detectors in these settings is another possible future research direction.

%\clearpage

%\section*{Author biographies}
%\parpic{\includegraphics[width=1in,clip,keepaspectratio]{Ravi.jpg}}
%\noindent {\bf Author Name} is the most famous author in her field.
%She is particularly interested in X and Y, and also dabbles in Z.

\bibliographystyle{IEEEtranJSAC}
\bibliography{IEEE_JSAC}

\end{document}